\documentclass[preprint, 11pt]{elsarticle}

\journal{Expert Systems with Applications}
\pdfoutput=1
\usepackage{booktabs}
\usepackage{graphicx}
\usepackage{hyperref}
\usepackage{array}
\usepackage{amsmath}
\usepackage{amssymb}
\usepackage[caption=false]{subfig}
\usepackage{multirow}
\usepackage{url}
\usepackage[linesnumbered, ruled]{algorithm2e}
\usepackage{tikz}
\usetikzlibrary{shapes,arrows,mindmap,decorations.pathmorphing,fit,positioning}

\newcommand{\dir}{\text{Dirichlet}}
\newcommand{\mult}{\text{Multinomial}}
\newcolumntype{M}[1]{>{\centering\arraybackslash}m{#1}}
\graphicspath{{./}}

\usetikzlibrary{shapes,arrows,positioning,decorations.pathreplacing, mindmap}
\tikzstyle{block} = [rectangle, draw, fill=blue!20, 
    text width=5em, text centered, rounded corners, minimum height=4em]
\tikzstyle{block1} = [rectangle, draw, fill=red!20, 
    text width=9em, text centered, rounded corners, minimum height=4em]
\tikzstyle{block2} = [rectangle, fill=white!20, 
    text width=5em, text centered, rounded corners, minimum height=4em]
\tikzstyle{line} = [draw, -latex']

\begin{document}

\begin{frontmatter}

\title{Enhanced Movie Content Similarity Based on Textual, Auditory and Visual Information}

\author[demo]{Konstantinos Bougiatiotis\corref{cor}}
\ead{bogas.ko@iit.demokritos.com}  %% follow author 1 immediately
\author[demo]{Theodoros Giannakopoulos}
\ead{tyiannak@gmail.com} %% If you want author 2 to have an email as well

\address[demo]{Institute of Informatics and Telecommunications,\\ National Center for Scientific Research ``Demokritos'', Athens 153 41, Greece }
\cortext[cor]{Corresponding Author: +30 693 1693844, Greece }

\begin{abstract}
In this paper we examine the ability of low-level multimodal features to extract movie similarity, in the context of a content-based movie recommendation approach. In particular, we demonstrate the extraction of multimodal representation models of movies, based on textual information from subtitles, as well as cues from the audio and visual channels. With regards to the textual domain, we emphasize our research in topic modeling of movies based on their subtitles, in order to extract topics that discriminate between movies. Regarding the visual domain, we focus on the extraction of semantically useful  features that model camera movements, colors and faces, while for the audio domain we adopt simple classification aggregates based on pretrained models. The three domains are combined with static metadata (e.g. directors, actors) to prove that the content-based movie similarity procedure can be enhanced with low-level multimodal information. In order to demonstrate the proposed content representation approach, we have built a small dataset of 160 widely known movies. We assert movie similarities, as propagated by the individual modalities and fusion models, in the form of recommendation rankings. Extensive experimentation proves that all three low-level modalities (text, audio and visual) \emph{boost the performance of a content-based recommendation system, compared to the typical metadata-based content representation, by more than $50\%$ relative increase}. To our knowledge, this is the first approach that utilizes a wide range of features from all involved modalities, in order to enhance the performance of the content similarity estimation, compared to the metadata-based approaches.

\end{abstract}


\end{frontmatter}

\section{Introduction}

In order to cope with the overwhelming amount of data available both online and offline, we are in dire need of recommendation systems, to browse through item collections and get meaningful recommendations. This is also the case when looking at \textit{motion pictures} in particular. There are several state-of-the-art systems providing movie recommendation services, most of which can be classified into either \textit{collaborative filtering} systems, such as \textit{MovieLens}\footnote{\url{https://movielens.org/}}, either \textit{content-based} systems, like \textit{jinni}\footnote{\url{http://www.jinni.com/}}, or \textit{hybrid} systems, as is \textit{IMDB}\footnote{\url{http://www.imdb.com/}}. More specifically, collaborative filtering systems are based on user preferences regarding the involved items, in order to make recommendations, while content-based systems use available descriptors of the movies to relate them with user preferences. However, all these systems rely on human-generated information, in order to create a corresponding representation and assess movie to movie similarity, not taking into account the raw content of the movie itself, but solely building upon annotations made by humans.

In this paper, we propose a method for representing movies, that is based directly on the movie's  audio, visual and textual content. Our vision is to incorporate knowledge regarding the way a movie ``sounds'' and ``look'' in the recommendation process. In this way, we differentiate from the related work (presented in the sequel), by providing latent representations of each movie, that could lead to explanatory results about the recommended movies, that take into consideration all ``aspects'' a movie.

The rest of this paper is organized as follows. Firstly, the related work is presented (Section \ref{sec:related}). Afterwards, the general workflow and details of the proposed method are explained (Section \ref{sec:prop}). We then present our data collection and ground truth generation methodology (Section \ref{sec:data_gt}). In the following section (Section \ref{sec:res}) the experimental results are presented and discussed. We close by drawing conclusions and outlining topics for further research (Section \ref{sec:fut}).

\section{Related Work}\label{sec:related}
Much research has been done on multimodal information extraction, specifically focused on video data sources. In the context of recommendation systems, there have been a number of studies \citep{Yang:2007, Mei:2011} focused on multimodal video recommendation, while other approaches are more application-specific, such as multimodal emotion classification \citep{tiwari:2016} or  
affective content analysis based on multimodal features \citep{Saran:2016}. For an in-depth overview of this field one can have a look at this survey \citep{brezeale2008automatic}. However, in the context of movie recommendation systems, the vast majority of existing approaches are based on collaborative knowledge or metadata \citep{miller2003movielens, wang2014improved}. 

The adoption of the \textit{multimedia} signal of the movies for indexing and/or recommendation, has been \textit{limited to particular applications}, such as emotion extraction \citep{malandrakis:emot, Kahou:2013} or violent content detection \citep{giannak:viol, nam:viol}. Other studies focus only on particular aspects of the movie, such as gender representation \citep{Guha:2015} or speaker clustering \citep{Kapsouras2017} using audiovisual features or movie topics generated from text \citep{Dupuy2017}. An application of deep convolutional networks is in \citep{farabet:2013} where the focus in on scene labeling from raw images. In addition, special focus has been given on video summarization, which is a rather important task that helps in extracting all the necessary information required from a video, without sacrificing much of the original informativeness. This task is often referred to as saliency estimation \citep{koutras2015predicting}. A state-of-the-art survey is reported in \citep{li:2013} with focus on video content analysis, representations and the possible applications of such endeavors.

Furthermore, audio-visual features have been adopted for movie genre classification \citep{rasheed2002movie}. In  \citep{deldjoo2016content}, a video recommendation system based on stylistic visual features is proposed, however, no other modalities are used. An extension of the previous system \citep{Deldjoo2017, Deldjoo:2017}, utilizes visual cues from trailers, as well as, human-generated tags. Another interesting recent work is in \citep{zhao2016matrix}, where movie recommendation  is done utilizing matrix factorization techniques on images  stemming from movie posters and frames. Regarding the audio domain, \citep{van:2013} use deep convolutional neural networks to predict latent factors from music audio signals and apply them in music recommendation. Deep learning frameworks have also been used in the textual domain of movies but not for recommendation purposes. Specifically, in \citep{serban:2015} they constructed a generative model for movie dialogues based on a hierarchical recurrent encoder decoder neural network.

Finally, a fresh idea is reported in \citep{wei:2016}, where a hybrid recommender system based on social movie networks and topic models is proposed with interesting results. Another hybrid recommender system is presented in \citep{singh:2011}. There, the authors focus on the fusion of collaborative filtering with sentiment classification of movie reviews to boost the final results. 

However, in this paper, we introduce the more ambitious objective of representing each movie \textit{directly from its raw multimodal content}. Our goal is to find correlations between similarity extracted from low-level feature modalities and high level association of those movies. This will lead us to innovative ways of defining movie similarity, explore \textit{latent} semantic knowledge from low-level cues and boost traditional information retrieval systems with information from heterogeneous content sources. The overall vision of adopting low-level modalities in content-based recommendation systems is two-fold:
\begin{itemize}
 \item to \textit{boost the performance of the recommendation} systems, by introducing new and diverse content descriptors that stem from low-level multimodal information
 \item to provide latent representations of the movies that can lead to \textit{knowledge discovery} and explanatory results about users' preferences (e.g. user's X preferences are highly influenced by the director's adopted techniques)
\end{itemize}

\begin{figure}[!htb]
    \centering
    \includegraphics[width = 1\textwidth]{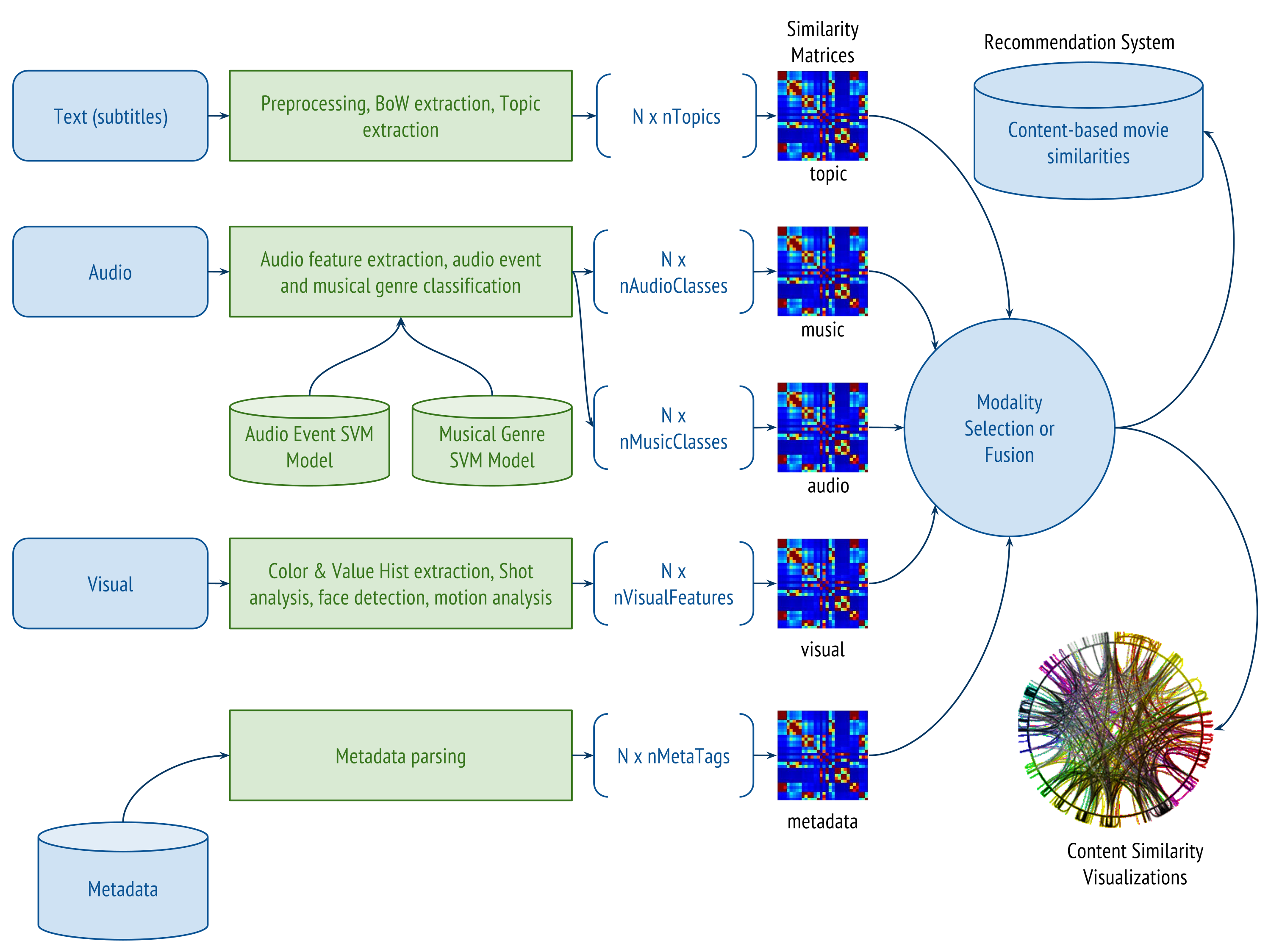}
    \caption{Workflow diagram of the proposed method. Different representations are extracted from each low-level modality (text, audio and visual), resulting in five individual content similarity matrices.}
    \label{fig:workflow}
\end{figure}

\section{Proposed Method}\label{sec:prop}
\subsection{General Workflow}
The overall scheme of the methodology described in the current work is presented in  Figure \ref{fig:workflow}. In summary, the following steps, with regard to the different modalities, are carried out :
\begin{itemize}
\item Text Analysis: Preprocessing, followed by the training of a topic model (through \textit{Latent Dirichlet Allocation}), of the subtitles for each movie. This textual analytics process is applied as the core component of the subtitle-based approach, in order to represent the textual content of the corresponding movies as vectors of topic weights. This results in a text representation matrix of $N$ rows ($N$ is the number of movies in the database) and $nTopics$ columns. 
\item Audio Analysis: two supervised audio models that represent the movie's content distribution to music and audio classes. This generic audio-based representation provides an aggregated projection of the types of sounds that appear in a movie. This twofold procedure results in two feature matrices for the whole dataset. Again, rows represent movies and columns represent the number of audio (or musical genre) classes respectively. An element of these matrices corresponds to the proportion of data classified to the respective audio (or musical genre) class. Both audio and musical genre matrices have 8 columns, since 8 audio classes and 8 musical genres are used in total.
\item Visual Analysis: features from the visual domain are extracted based on the distribution of colors, camera movement, existence of faces in the scenes and shot lengths. This results in a feature matrix of visual characteristics. Rows represent movies and columns visual features. As explained in the sequel, the number of columns of this matrix is 208.
\item Metadata Analysis: Metadata information about each movies' cast, director and genre are parsed into categorical feature vectors, in order to evaluate the ability of these handcrafted attributes to extract similarity measures between movies, and  combine this type of information with subtitles and audio-visual content similarity. Using metadata is not the core idea of this paper, however we adopt their usage in order to demonstrate the ability of the low-level multimodal features to boost the performance of the content similarity procedure.
\item Content Similarity Fusion: Fusing the similarity matrices that were generated through the previous steps, we yield multimodal similarity measures between movies. In the context of this work, we have focused on a simple and straightforward fusion approach that applies weighted averaging on the individual content similarity matrices.

\end{itemize}
Our goal in the context of this work is to prove that these low-level modalities can improve the performance of the metadata-based content similarity estimation, when combined.

\subsection{Subtitles Analysis}
\subsubsection{Preprocessing}
We start by applying a series of  essential preprocessing steps on each subtitles' document, since these documents are .\textit{srt} files in our dataset, filled with unwanted information, such as timestamps and markup elements. Moreover, we also want to filter out noisy data and non-informative words that do not add to the distinctiveness of the documents. In particular, the following transformations are applied in all documents:
\begin{itemize}
\item \textbf{Regular expressions removal}: remove mark-up elements, timestamps and anything not content-related
\item \textbf{Tokenization}: case-folding and splitting up the textual strings to words using whitespaces
\item \textbf{Lemmatization} unify terms stemming from the same lemma having differences due to inflectional morphologies. The lemmatizer used to this end is based on the \textit{WordNet} database \citep{fellbaum:wordnet}.
\end{itemize}

Afterwards, we move on to \textbf{word filtering}. Firstly, common, movie-domain specific and subtitle related stopwords  are removed. These are mainly common words, taken from the \textit{nltk stopwords corpus}\footnote{\url{http://www.nltk.org/nltk_data/}}, such as \textit{``I''}, \textit{``it''}, \textit{``and''} etc. that do not offer any additional information to the document. We also manually selected words that are common ground in the subtitles or contain errors such as \textit{``aint''}, \textit{``Ill''}, \textit{``theres''} and \textit{``yeah-yeap-yess''}. We also remove words which provide low information for each document. These are words with low intra-document and high inter-document frequency. The core idea is that words appearing only a few times in each document or words appearing in most of the documents in our collection are not useful in order to differentiate them. This is also useful for trimming the total vocabulary size, thus cutting down the dimensions for the representation space of the documents countering the problems of sparsity and fragmentation of vector-term space.

After the aforementioned processes, each movie can be thought of as a \textit{bag of words} (\textit{BoW}). This allows us to model each document as a vector in this term space, with values in each cell denoting the number of occurrences of the corresponding word.

\subsubsection{Content representations}
There are many ways to use the aforementioned BoW vectors in information retrieval applications. In the context of our work we will mainly focus on \textit{Latent Semantic Indexing} (\textit{LSI}) and \textit{Latent Dirichlet Allocation} (\textit{LDA}). However, we first describe the \textit{term frequency-inverse document frequency} (\textit{tf-idf}) weighting scheme, which is used mainly for benchmarking purposes as the most easily implement methodology among the ones mentioned.

\paragraph{Term frequency-Inverse Document Frequency (tf-idf)}
Tf-idf is a weighting scheme, where the words in the BoW representation of the documents are allocated a weight denoting the importance of the word for the specific document \citep{salton:information}. The weight is computed based on two different factors. The first is the term frequency in the document and denotes the importance of the word for the specific document \citep{luhn:1957}, while the second is a factor inversely proportional to the frequency of the term over the whole collection of the documents \citep{jones:1972}. The resulting weight, fusing  those two sources of information, is calculated as:

\begin{equation}
tf\text{-}idf_{i, d} = tf_{i, d}\times idf_i = tf_{i, d}\times \log_2 {\frac{N}{n_i}}
\end{equation}

where $tf_{i, d}$ is the absolute frequency of term $i$ in document $d$, $N$ the number of documents and $n_i$ the number of documents in our collection in which term $i$ appears.

\paragraph{Latent Semantic Indexing (LSI)} Although \textit{tf-idf} is a powerful tool, there are more sophisticated methods that mainly deal with the problems of sparsity and dimensionality of the document-term representation used in the previous scheme. Moreover, they also address the problems of synonymy and polysemy. One of the most widely used methods is \textit{Latent Semantic Indexing} (\textit{LSI}) (or \textit{Latent Semantic Analysis} (\textit{LSA}) as referred in other domains) \citep{deerwester:1990}. The core idea behind this methodology is that instead of projecting documents in the multidimensional term space, we can project both the terms and the documents in a much lower dimensionality space, whose axes represent concepts that essentially group words together. These axes are the Principal Components from Principal Components Analysis \citep{pearson:1901} that exhibit the greatest variation and are propelled from the co-occurence of words in the documents. In this way two documents can have high similarity in the latent semantic space without containing the same words, leading to interesting results in terms of information retrieval.

\begin{figure}[ht]
\centering
\includegraphics[width = 1\textwidth]{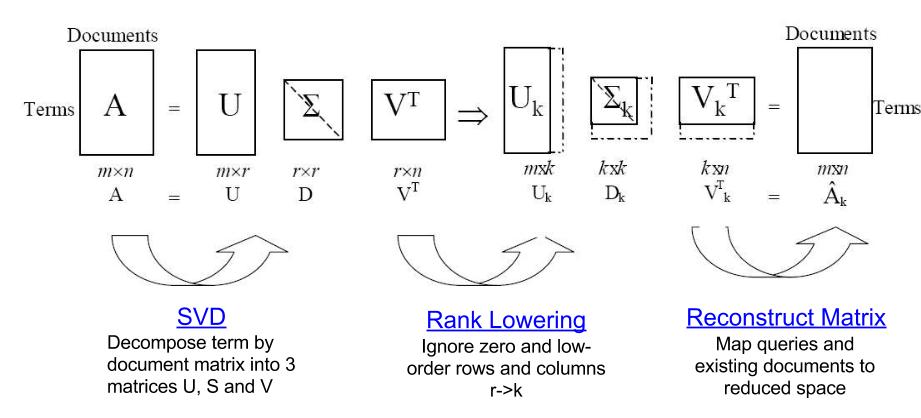}
\caption{Singular value decomposition followed by rank lowering for latent semantic indexing}
\label{fig:lsi}
\end{figure}

LSI is essentially a two-step method as illustrated in Figure \ref{fig:lsi}\footnote{Source:  https://liqiangguo.files.wordpress.com/2011/06/lsi2.pdf} that uses a low-rank approximation of the document-term matrix created from the term-vector space projections. Firstly, \textit{singular value decomposition} \textit{(SVD)} is applied on the document-term matrix, where the newly created eigenvectors represent the concepts in the latent space. Secondly, lower order columns are ignored and only the first k principal concepts of the eigenvalues/eigenvectors matrices, reducing the dimensionality of the representation, thus cutting down noise in the latent space, resulting in a richer word relationship structure that reveals latent semantics present in the collection. Now, using those lower-dimensionality matrices we can map documents (movies in our case) in the latent concept space and calculate similarity between movies in this richer representation space. 

\begin{figure}[ht]
\centering
\includegraphics[width = 1\textwidth]{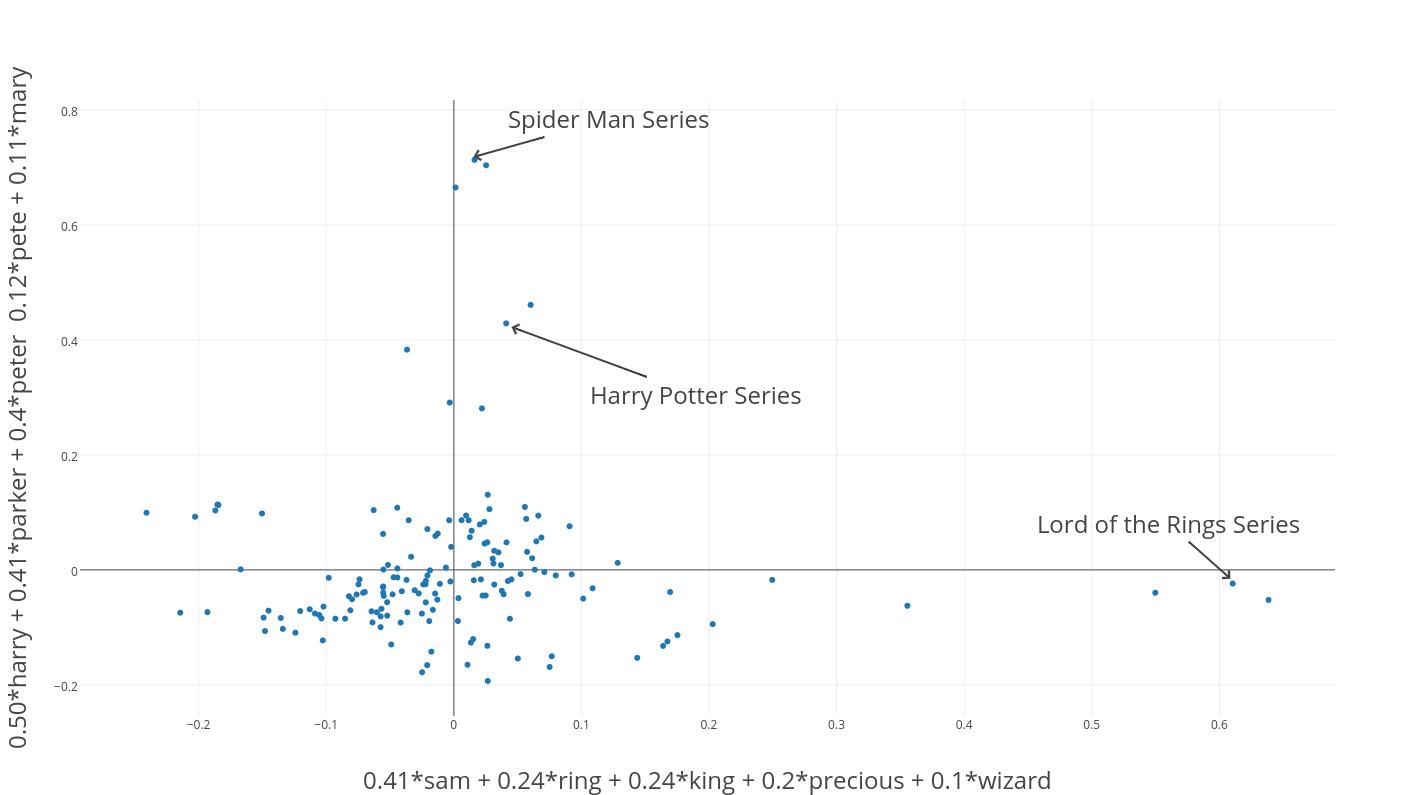}
\caption{Projection in a 2d-concept space of the movies. The x-axis concept is related to the Lord of the Rings trilogy, while the y-axis is related to the Spider Man Movies.}
\label{fig:lsi2}
\end{figure}

Note that we use the LSI method after implementing the tf-idf transform on the document-word matrix of our collection, but this is not mandatory, since other initial document representations could also be used instead (e.g. simple word counts). Also, the order of dimensionality reduction that LSI imposes on the vector space model (the number of principal concepts to keep) is a user-defined parameter. After, extensive experimentation in our setup we selected $T=55$ concepts to be the optimal value. In order to visualize the projection of the movies in the concept space, we illustrate in Figure \ref{fig:lsi2} a 2-d reconstructed example. The x-axis concept is related to the \textit{Lord of the Rings} trilogy, with important words as those shown in the caption of the axis, while the y-axis is related to the Spider Man Movies. The words shown in the axes are the coefficients of the most influential words for this specific eigenvector/concept of the LSI method. The top cluster of movies are the \textit{Spider Man} movies, while the utmost right cluster of movies is correspondingly the Lord of the Rings movies, as expected. Notice however, the mid level center cluster with movies belonging to the \textit{Harry Potter} series. They have large values in the y-axis concept because of the high influence of the word ``harry'' (i.e. ``harry osborn''  is the name of a character in \textit{Spider Man}) in the y-axis concept, among other words.

\paragraph{Latent Dirichlet Allocation (LDA)} In order to deal with the shortcomings of LSI, like the fact that LSI does not take into account that the scores in the document-term matrix come from term frequencies and the resulting eigenvectors may lead to negative coefficients in the concept space, we also implement a topic modeling algorithm, namely \textit{Latent Dirichlet Allocation} (\textit{LDA}) \citep{blei:lda}. This is a probabilistic generative model structured upon the idea that all documents (movies) can be thought of as a mixture of specific topics.Each movie exhibits those topics in different proportions, so alike movies tend to exhibit more or less the same topics. Each topic is a distribution over the words in the vocabulary of our collection. LDA is a generative process, meaning that each document in our collection can  be created through a structured process, given a set of hidden variables. Specifically, Algorithm \ref{algo:LDA} describes they way in which the documents in our collection are generated.

\begin{algorithm}[!htb]
\SetAlgoLined
\For{each topic $\beta_k, k:1..K$ }{
    Choose $\beta_k \sim \dir(\eta)$ \tcp*[r]{\small A distr. over words for the topics} 
}
 \For{each movie $d_d$ in our collection}{
  Choose $\theta_d \sim \dir(\alpha) $ \tcp*[r]{ \small A distr. over topics for the document} \
  \For{word $w_n$ in $d$}{Choose a topic the word belongs to$z_{d,n} \sim \mult(\theta_d)$\; Choose a word $w_{d,n}$ from $p(w_{d,n} | z_{d,n},\beta_{z_{d,n}}) \sim \mult(\beta_{z_{d,n}})$
      }
  }
 \caption{Generative process of LDA}
 \label{algo:LDA}
\end{algorithm}

The aforementioned procedure is based on two hidden variables. Firstly, the topic distributions over words $\beta_k, \forall k$ topics and the distribution of documents over topics $\theta_d, \forall d$ documents. Using the available documents, our goal is to infer the posterior distribution of these hidden variables given the observed ones, namely the document-words matrix. The variables needed are the number of topics $K$ that exist in our collection and the hyperparameters of the Dirichlet distributions $\eta, \alpha$. These parameters control the sparsity of the topic-word relations and document-topic distributions and are mainly calculated heuristically \citep{griffiths:2004}. Maximum a posteriori estimation is intractable for this model \citep{dickey:hypergeo}, however there are many variatonal and  sampling methods for approximation of the wanted posterior. In our case, we used a \textit{Collapsed Gibbs Sampling} version of the algorithm \citep{mccallum:mallet} and more specifically its implementation in the Gensim library \citep{rehurek:gensim}. We defined $K=55$ topics after experimenting with the documents in our collection, while hyperparameters $\eta, \alpha$ are both optimized during fitting of the model \citep{minka:hyper}.

\begin{figure}[ht]
\centering
\includegraphics[width = 1\textwidth]{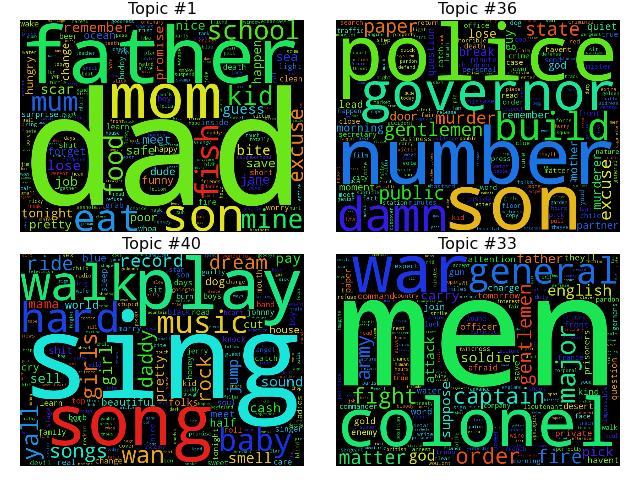}
\caption{Word Clouds examples for 4 Topics}
\label{fig:clouds}
\end{figure}

In order to give a qualitative example of the generated topics we illustrate some of them as word clouds in Figure \ref{fig:clouds}. The size of each word is proportional to the importance of the word for this topic. If we observe the resulting topics, we can see that they are well formulated and coherent. For example, the top left topic is highlighted by words such as \textit{dad, father, mom, son, school}, defining a family related topic while the bottom right exhibits mainly words like \textit{men, colonel, war, general}, defining a war related topic. This semantically concise and friendly way of representing the topics existing in our collection is another reason why LDA is sometimes opted in favor of LSI.

\begin{figure}
\centering
\begin{tikzpicture}[grow cyclic, text width=2.1cm, align=flush center, 
    level 1/.style={level distance=0.3\textwidth,sibling angle=270, counterclockwise from=-45},
    level 2/.style={level distance=0.25\textwidth,sibling angle=30, counterclockwise from=180}]
\node{Topic Model}
    child { node {hope, prison, security, escape, state}
           child[clockwise from=180]{ node {\includegraphics[width=0.45\textwidth]{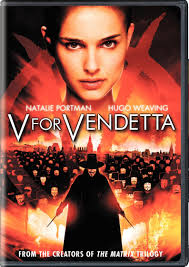}}}
        child { node {\includegraphics[width=0.45\textwidth]{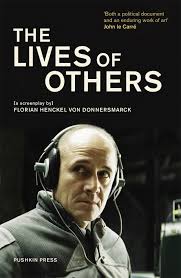}}}
        child { node {\includegraphics[width=0.45\textwidth]{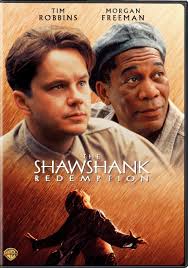}}}
        child { node {\includegraphics[width=0.45\textwidth]{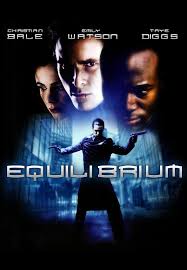}}}
   }
    child { node {school, dad, mom, house, parent} 
    		child{ node {\includegraphics[width=0.45\textwidth]{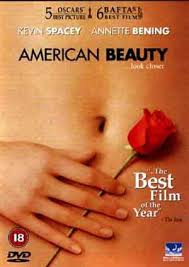}}}
        child { node {\includegraphics[width=0.45\textwidth]{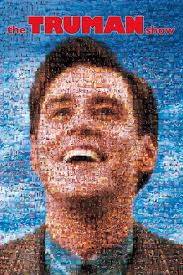}}}
        child { node {\includegraphics[width=0.45\textwidth]{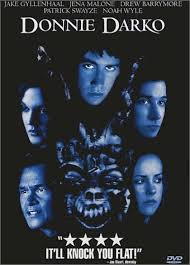}}}
        child { node {\includegraphics[width=0.45\textwidth]{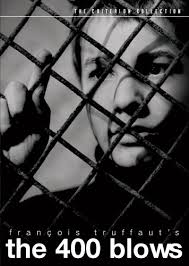}}}
    }
;
\end{tikzpicture}
 \caption{Clustering of movies based on specific topics.}
 \label{fig:topic_recomm}
\end{figure}

Moreover, we demonstrate in Figure \ref{fig:topic_recomm} the usefulness of the learned topic model in clustering certain movies together based on their relevance through specific topics. Here, we showcase the most influential movies tethered to two specific topics, as generated from our collection. One topic is from the word cloud in the previous figure, with words about family, school etc. and the other one is related to imprisonment, security and the state. As you can see for the figure, \textit{American Beauty, Donnie Darko, The 4oo blows} and \textit{Truman Show} have been clustered together as movies focused on the first topic, while \textit{V for Vendetta, The Lives of Others, Shawshank Redemption} and \textit{Equilibrium} have been brought together nicely as co-thematic movies about the latter topic. 

This higher level of representation and thematic browsing of the movies is an invaluable tool, in order to get content-generated recommendations that conventional recommendation systems can't offer.

\subsection{Audio Analysis}
The audio signal  is a very important channel of information with regards to a movie's content: music tracks, musical  background themes, sound effects, speech, acoustic events, they all play a vital role in forming the movie's ``style''. Therefore, a content representation approach should also take into account these aspects of information. Towards this end, in the presented method we have extracted two types of information: (a) music-genre statistics and (b) audio event statistics. 

In particular, we have trained two separate supervised models using Support Vector Machines, in order to classify all movie audio segments to a set of predefined classes related either to audio events or musical genres. To this end, the \textit{pyAudioAnalysis} \citep{giannakopoulos2015pyaudioanalysis}  library has been used to extract audio features both in a short-term and in a mid-term basis. The movie audio stream is split to non-overlapping segments of 2 seconds. For each mid-term segment, a set of mid-term feature statistics (described in \citep{giannakopoulos2015pyaudioanalysis}) is extracted to represent its content.

This feature vector is fed as input to the audio event classifier, which decides for the respective class label. The adopted classes for this task are: music, speech, 3 types of environmental sounds (low energy background noise, abrupt sounds and constant high energy sounds), gunshots-explosions, human fights and screams (8 classes in overall). Furthermore, each segment classified as ``music'' is also fed as input to a musical genre classifier, which decides among the following classes: jazz, classical, country, blues, electronic, rap , reggae and rock. The result of this process is a sequence of music-genres and a sequence of audio events. Note that, in order to train the two classifiers a separate and independent dataset has been annotated.  The final representation that corresponds to the whole movie is provided by two vectors that represent the proportions of each musical-genre or audio event class.

\begin{figure}[ht]
    \centering
    \includegraphics[width = 0.9\textwidth]{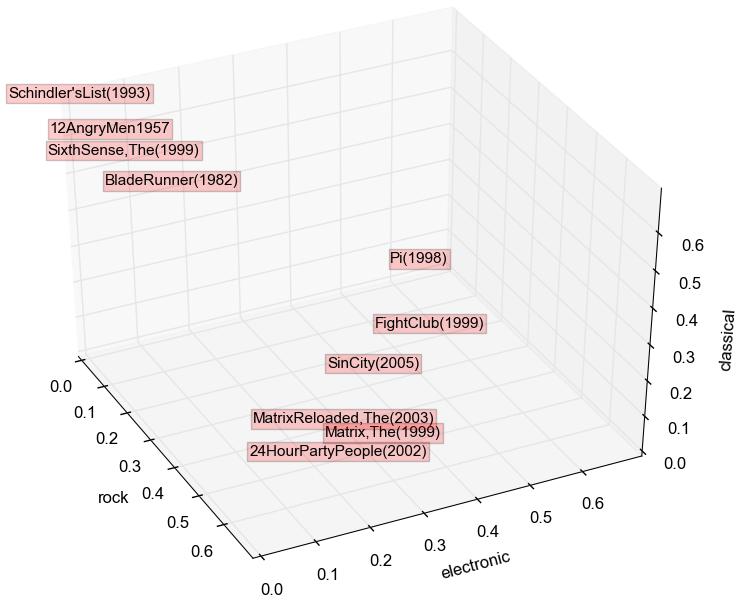}
    \caption{Examples of 10 movies distributed in 3 musical genres. The values in the $[0,1]$ range correspond to the proportion of musical segments classified to the respective musical genre class. }
    \label{fig:demoMusic}
\end{figure}

Figure \ref{fig:demoMusic} presents an example of three musical-genre-related features for 10 movies. The three features correspond to the proportion of music segments classified as ``rock'', ``electronic'' or ``classical''. Three obvious ``clusters'' can be observed: movies with classical music themes (e.g. \textit{Schindler's List}), movies with almost equal distributions of electronic and rock music segments (e.g. \textit{The Matrix} and \textit{24 Hour Party People}) and two movies (\textit{Pi}, \textit{Fight Club}) that are mostly related to electronic music. 

\subsection{Visual Analysis}
Visual information contains the major characteristics of a movie regarding its filming techniques and its type of involved actions, so it can be considered as a richer domain compared to the audio medium. Our goal with regards to the visual channel, in the context of the presented research effort, is to extract low-level visual features that express latent semantic attributes that discriminate between different cinematic techniques and film contents. 

\begin{table}[!htb]
      \centering
        \begin{tabular}{p{0.14\textwidth}p{0.08\textwidth}p{0.16\textwidth}p{0.48\textwidth}}\hline
        Category & Indices & Name & Description \\ \hline
        \multirow{6}{0.15\textwidth}[-2em]{Color and illumination} & 0-7      & R Hist & Histogram of the red coordinate (RGB) \\
                              & 8-15     & G Hist & Histogram of the green coordinate (RGB) \\
                              & 16-23    & B Hist & Histogram of the blue coordinate (RGB) \\
                              & 24-31    & V Hist & Histogram of the grayscale values of the frame\\
                              & 32-36    & RGB ratio Hist & Histogram of the rgb-ratio color\\                 
                              & 37-44    & S Hist & Histogram of the saturation coordinate of the HSV color space\\\hline
        \multirow{2}{0.15\textwidth}[-1em]{Faces} & 46       & NFaces   & Number of detected faces in the frame \\
                              & 47       & PerFaces & Average ratio of each face bounding box's area to the whole frame area \\\hline
        \multirow{5}{0.15\textwidth}[-2em]{Motion}       & 45       & Gray Diff & Mean absolute difference between two successive frames  \\    
                              & 48       & Tilt-Pan Measure & A flow-based feature that measures tilting and panning movements \\
                              & 49       & Flow Mean Mag & Average magnitude of the flow vectors \\
                              & 50       & Flow Std Mag & Standard deviation of the magnitudes of the flow vectors \\\hline
        Shot-related         & 51       & Shot & Estimated duration of the shot that contains the current frame \\ \hline
        \end{tabular}
        \caption{List of adopted frame-wise visual features. The final movie representation is a vector of $52 \times 4 = 208$ feature statistics.} \label{tab:visualFeatures}
\end{table}

Table \ref{tab:visualFeatures} presents the list of features extracted. These features are extracted on a frame basis, i.e. for each frame of the movie. For reducing computational complexity, all frames are re-sized to a fixed width of 500 pixels. In addition, we process 2 frames per second, since experiments have shown that this is an adequate rate for the adopted features. This process leads to a $nFrames \times 52$ feature matrix, where rows correspond to frames and columns to visual features. The final feature vector that represents the whole film results from four statistics applied on the aforementioned feature sequences. In particular, the following statistics are computed for each feature sequence:
\begin{enumerate}
\item average value $\mu$
\item standard deviation $\sigma^2$
\item $\frac{\sigma^2}{\mu}$ ratio
\item average value of the top $10\%$ highest feature values
\end{enumerate}

After the statistics calculation, each movie is represented by a 208 ($52$ features $\times$ $4$ statistics) feature vector. In the rest of this section, we describe the adopted visual features along with examples that demonstrate their ability to discriminate between cinematic attributes and correspond to high-level similarities of movies. 

\subsubsection{Color and Illumination}

Adopted colors and color effects play a vital role in the  director's effort to enhance the mood or to punctuate a dramatic tone in the movie. Color and illumination differentiations in cinematic movies are either due to the illustrated subjects and locations or to an artistic process. In many cases, digital color correction is deliberately applied to convey a particular artistic perspective or tone. 

\begin{figure}[!htb]
\centering
        \subfloat{\centering\includegraphics[width=0.20\textwidth]{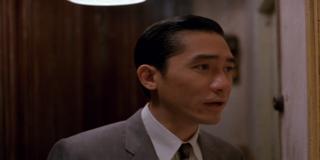}}
        \subfloat{\centering\includegraphics[width=0.20\textwidth]{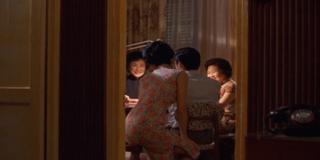}}
        \subfloat{\centering\includegraphics[width=0.20\textwidth]{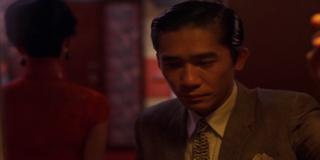}}
        \subfloat{\centering\includegraphics[width=0.20\textwidth]{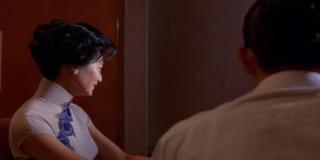}}
        \vspace{-1.2em}
        \subfloat{\centering\includegraphics[width=0.20\textwidth]{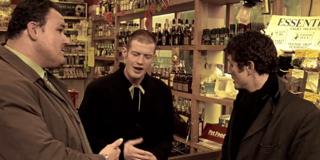}}
        \subfloat{\centering\includegraphics[width=0.20\textwidth]{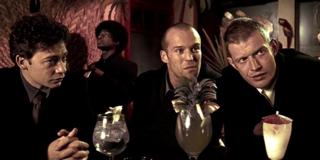}}
        \subfloat{\centering\includegraphics[width=0.20\textwidth]{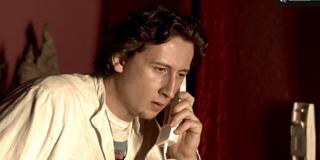}}
        \subfloat{\centering\includegraphics[width=0.20\textwidth]{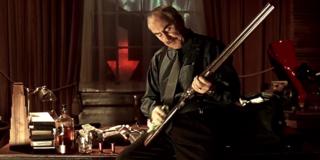}}
        \vspace{-1.2em}
        \subfloat{\centering\includegraphics[width=0.20\textwidth]{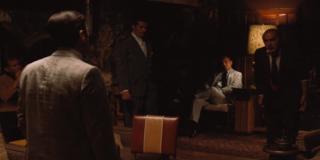}}
        \subfloat{\centering\includegraphics[width=0.20\textwidth]{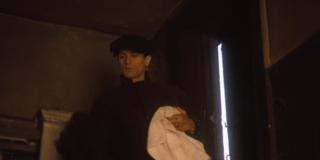}}
        \subfloat{\centering\includegraphics[width=0.20\textwidth]{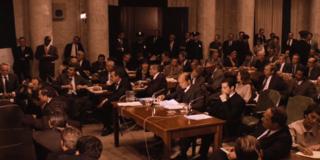}}
        \subfloat{\centering\includegraphics[width=0.20\textwidth]{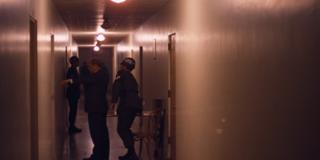}}
        \vspace{-1.2em}
        \subfloat{\centering\includegraphics[width=0.20\textwidth]{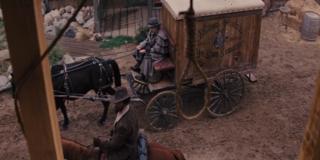}}
        \subfloat{\centering\includegraphics[width=0.20\textwidth]{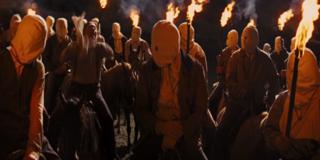}}
        \subfloat{\centering\includegraphics[width=0.20\textwidth]{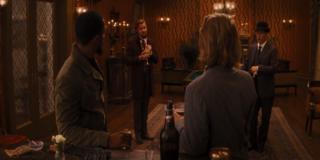}}
        \subfloat{\centering\includegraphics[width=0.20\textwidth]{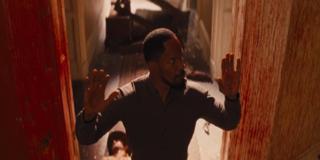}}
        \vspace{-1.2em}
        \subfloat{\centering\includegraphics[width=0.20\textwidth]{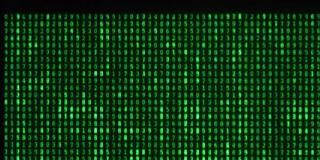}}
        \subfloat{\centering\includegraphics[width=0.20\textwidth]{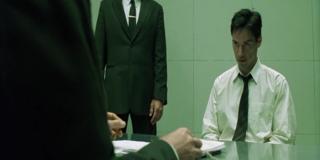}}
        \subfloat{\centering\includegraphics[width=0.20\textwidth]{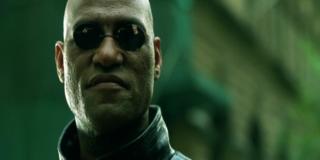}}
        \subfloat{\centering\includegraphics[width=0.20\textwidth]{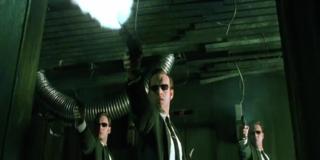}}
        \vspace{-1.2em}
        \subfloat{\centering\includegraphics[width=0.20\textwidth]{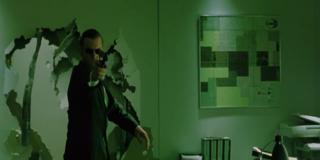}}
        \subfloat{\centering\includegraphics[width=0.20\textwidth]{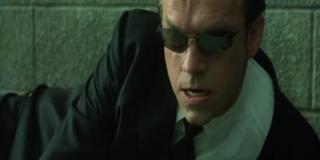}}
        \subfloat{\centering\includegraphics[width=0.20\textwidth]{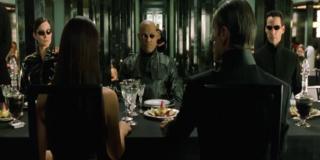}}
        \subfloat{\centering\includegraphics[width=0.20\textwidth]{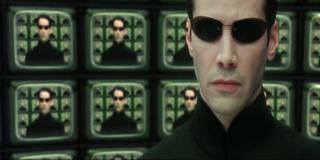}}
        \vspace{-1.2em}
        \subfloat{\centering\includegraphics[width=0.20\textwidth]{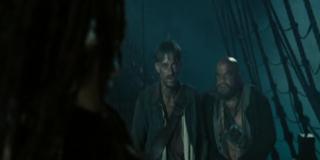}}
        \subfloat{\centering\includegraphics[width=0.20\textwidth]{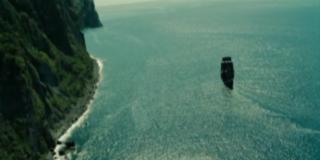}}
        \subfloat{\centering\includegraphics[width=0.20\textwidth]{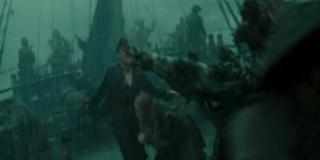}}
        \subfloat{\centering\includegraphics[width=0.20\textwidth]{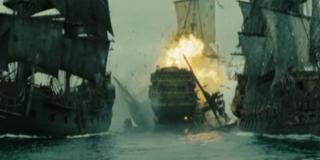}}
        \vspace{-1.2em}
        \subfloat{\centering\includegraphics[width=0.20\textwidth]{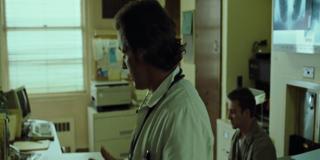}}
        \subfloat{\centering\includegraphics[width=0.20\textwidth]{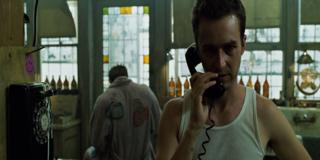}}
        \subfloat{\centering\includegraphics[width=0.20\textwidth]{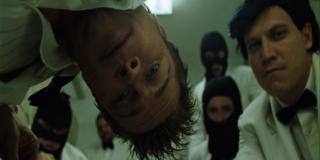}}
        \subfloat{\centering\includegraphics[width=0.20\textwidth]{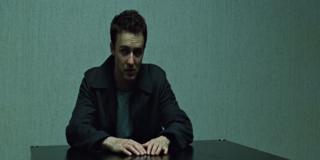}}
        \vspace{-1.2em}
        \subfloat{\centering\includegraphics[width=0.20\textwidth]{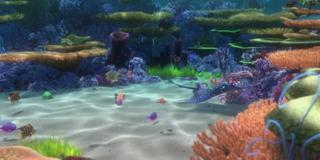}}
        \subfloat{\centering\includegraphics[width=0.20\textwidth]{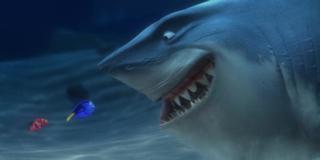}}
        \subfloat{\centering\includegraphics[width=0.20\textwidth]{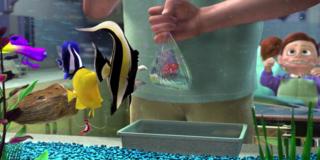}}
        \subfloat{\centering\includegraphics[width=0.20\textwidth]{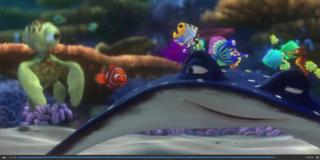}}
        \vspace{-1.2em}
        \subfloat{\centering\includegraphics[width=0.20\textwidth]{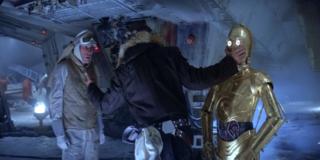}}
        \subfloat{\centering\includegraphics[width=0.20\textwidth]{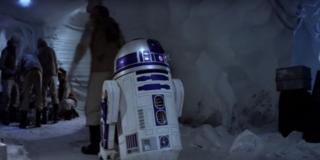}}
        \subfloat{\centering\includegraphics[width=0.20\textwidth]{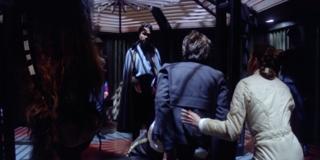}}
        \subfloat{\centering\includegraphics[width=0.20\textwidth]{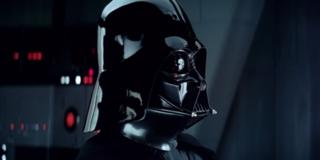}}
        \vspace{-1.2em}
        \subfloat{\centering\includegraphics[width=0.20\textwidth]{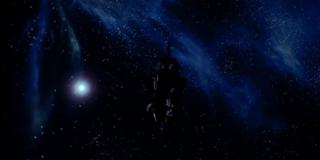}}
        \subfloat{\centering\includegraphics[width=0.20\textwidth]{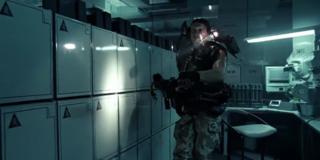}}
        \subfloat{\centering\includegraphics[width=0.20\textwidth]{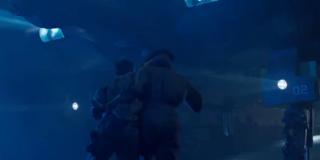}}
        \subfloat{\centering\includegraphics[width=0.20\textwidth]{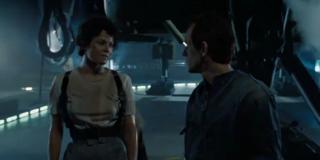}}
        \vspace{-1.2em}
        \subfloat{\centering\includegraphics[width=0.20\textwidth]{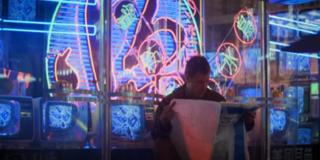}}
        \subfloat{\centering\includegraphics[width=0.20\textwidth]{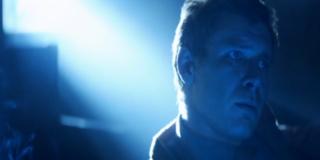}}
        \subfloat{\centering\includegraphics[width=0.20\textwidth]{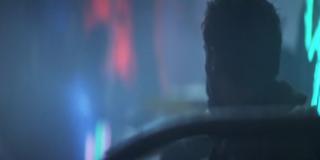}}
        \subfloat{\centering\includegraphics[width=0.20\textwidth]{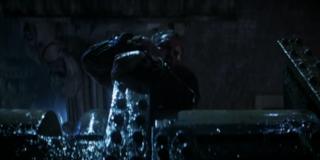}}

\caption{Screenshots from four movies per most dominant color coefficient. 
Red was most dominant in the first set of movies, green was most dominant in movies 5-8 and blue was most dominant in the final set of movies. 
} \label{fig:colorDemo1}
\end{figure}

In order to model color and illumination the following visual features are extracted:
\begin{itemize}
\item \textit{RGB histograms}: for every color coordinate (red, green and blue) an 8-bin histogram is computed
\item \textit{Value histogram}: an 8-bin histogram is computed on the grayscale values of each frame, in order to model the distribution of the movie's illumination
\item \textit{RGB ratio}: a simple measure of each frame's color saturation is extracted as the ratio of the maximum RGB value to the average RGB value (at each pixel). Then, a 5-bin histogram of this new image is extracted as the final feature (note: 5 bins instead of 8 because the RGB ratio is thresholded for significantly low and  high values, for the sake of normalization)
\item \textit{Saturation histogram}: another color saturation feature set is extracted as the histogram of the S coordinate of the HSV color space.
\end{itemize}

\begin{figure}[!htb]
\centering
        \subfloat{\centering\includegraphics[width=0.20\textwidth]{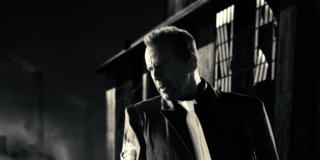}}
        \subfloat{\centering\includegraphics[width=0.20\textwidth]{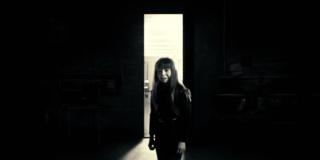}}
        \subfloat{\centering\includegraphics[width=0.20\textwidth]{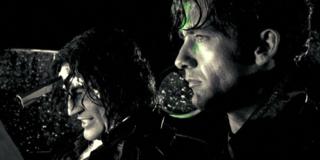}}
        \subfloat{\centering\includegraphics[width=0.20\textwidth]{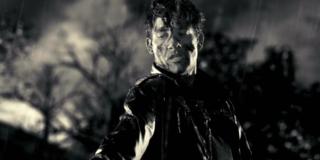}}
        \vspace{-1.2em}
        \subfloat{\centering\includegraphics[width=0.20\textwidth]{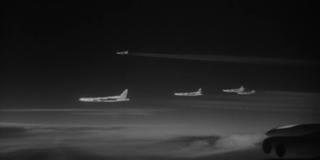}}
        \subfloat{\centering\includegraphics[width=0.20\textwidth]{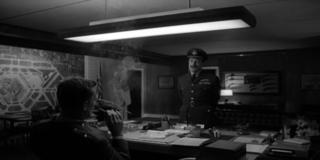}}
        \subfloat{\centering\includegraphics[width=0.20\textwidth]{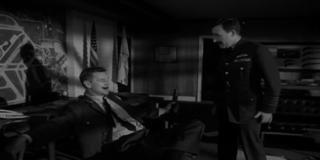}}
        \subfloat{\centering\includegraphics[width=0.20\textwidth]{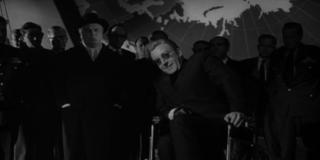}}
        \vspace{-1.2em}
        \subfloat{\centering\includegraphics[width=0.20\textwidth]{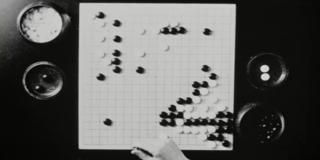}}
        \subfloat{\centering\includegraphics[width=0.20\textwidth]{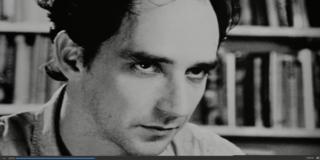}}
        \subfloat{\centering\includegraphics[width=0.20\textwidth]{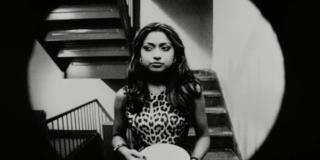}}
        \subfloat{\centering\includegraphics[width=0.20\textwidth]{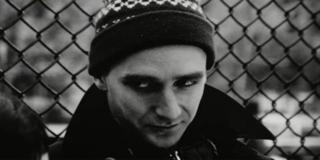}}   
\caption{Two of the darkest movies (\textit{Sin City} and \textit{Dr. Strangelove}) and the lightest movie (\textit{Pi}). \textit{Pi} is also the movie with the highest illuminance diversity. 
} \label{fig:colorDemo2}
\end{figure}

Figure \ref{fig:colorDemo1} presents screenshots from typical movies with dominant RGB coordinates. Each row corresponds to a different movie. Rows 1-4 correspond to movies which have red as a dominant color (\textit{In the Mood For Love}, \textit{Lock, Stock and Two Smoking Barrels}, \textit{Godfather II} and \textit{Django Unchained}), rows 5-8 to movies with green (\textit{The Matrix}, \textit{The Matrix Reloaded},\textit{ Pirates of the Caribbean: At World's End} and \textit{Fight Club}) and rows 9-12 to blue movies (\textit{Finding Nemo}, \textit{Star Wars Episode V - The Empire Strikes Back}, \textit{Aliens} and \textit{Blade Runner}). In all cases, the selection of color corresponds to a intentional choice made by the producers to express either meaning (e.g. red is usually adopted to express violence, guilt and sin), mood or even a particular era (warm colors are adopted in many movies set in the 60s and 70s). In some cases, dominant colors express a particular plot concept, for example in \textit{The Matrix} sequel, the directors' color pallet choice refers to the monochrome monitors used in early computing, and is used to discriminate between the ``real'' and the Matrix world. 

\begin{figure}[!tpb]
\centering

        \subfloat{\centering\includegraphics[width=0.20\textwidth]{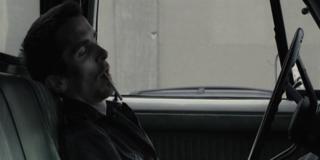}}
        \subfloat{\centering\includegraphics[width=0.20\textwidth]{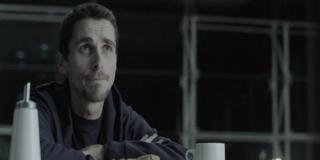}}
        \subfloat{\centering\includegraphics[width=0.20\textwidth]{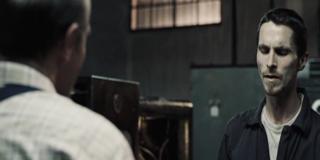}}
        \subfloat{\centering\includegraphics[width=0.20\textwidth]{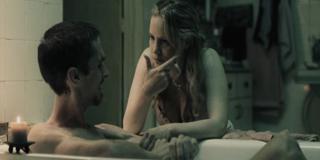}}
        \vspace{-1.2em}
        \subfloat{\centering\includegraphics[width=0.20\textwidth]{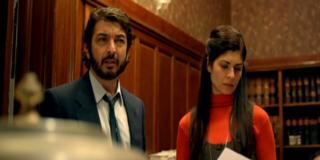}}
        \subfloat{\centering\includegraphics[width=0.20\textwidth]{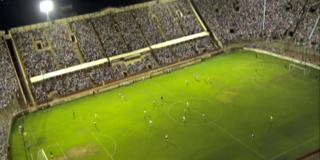}}
        \subfloat{\centering\includegraphics[width=0.20\textwidth]{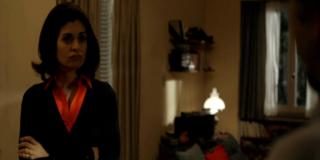}}
        \subfloat{\centering\includegraphics[width=0.20\textwidth]{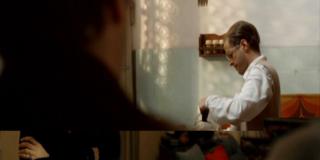}}
\caption{Movies with the lowest (\textit{Machinist}) and highest (\textit{The Secrets in their Eyes}) saturation. Black and white movies are obviously excluded. 
} \label{fig:colorDemo3}
\end{figure}

In Figure \ref{fig:colorDemo2} two movies with the highest average illumination and one movie with the lowest illumination are presented. The particular examples' values are directly extracted from the first bin of the value histogram described above. \textit{Pi}, which is the lightest movie has also been found to have the highest illumination diversity (i.e. the ratio of the darkest bin to the lightest bin). Indeed, the frames from this movie have a characteristic black and white range of gray values, quite close to a binarized image. Finally, Figure \ref{fig:colorDemo3} shows the less and most saturated movies. \textit{Machinist} is indeed a typical example of a extremely desaturated movie.

\subsubsection{Motion}
Along with colors, motion is the most important visual characteristic of a film and differentiates between different movie genres and filming techniques. It can be either due to the subject's movement, therefore depends on the particular type of recorded action, or due to the camera movement methodology. In the context of this work we have implemented the following motion-related features:

\textit{Frame difference between two successive frames.}
This feature is simply computed as a mean absolute distance between the values of two successive frames.

\paragraph{Flow-based features} Optical flow \citep{horn1981determining} has been widely used in motion estimation and video encoding. In this work, we estimate flow vectors using a sparse iterative version of the Lucas-Kanade optical flow in pyramids \citep{bouguet2001pyramidal}. After estimating the flow vectors, we move on to detect typical camera movements. In particular, we focus on the following 
 cinematographic techniques with regards to camera movement:
\begin{itemize}
    \item pan: the camera is rotated horizontally from a fixed position
    \item tilt: the camera is rotated vertically from a fixed position
    \item pedestal: the camera is moving on the vertical axis, without change in the horizontal axis
    \item truck: the camera is moving left or right (i.e. on the horizontal axis), without change in its perpendicular location
\end{itemize}

In all four methods, the perceived motion of the scene is similar: all points seem like moving in the same direction. Therefore, we expect that the flow vectors computed over scenes that are characterized by such camera movements will share (almost) the same angle. Based on that idea, we compute the following features:
\begin{itemize}
    \item pan-tilt-pedestal-truck (PTPT) confidence  movement: for each frame if $F_i, i=1,\ldots,N$ are the magnitudes and $\phi_i, i=1,\ldots,N$ are the angles of the $N$ flow vectors, we compute the following measure: $\frac{\sum_{i=0}^{N} F_i }{\sum_{i=0}^{N} [\Delta(\phi_i, \bar{\phi})]^2}$
    which is maximized for high magnitude values (therefore high motion velocities) and low deviation of the angles (which corresponds to near-parallel flow vectors). Note that $\Delta(x, y)$ is the angle difference between angles $x$ and $y$, and $\bar{\phi}$ is the mean value of $\phi$.
    \item the average value of the flow vector magnitudes: $\frac{1}{N}\sum_{i=0}^{N} F_i$
    \item the deviation of the flow vector angles $\frac{1}{N} \sum_{i=0}^{N}[\Delta(\phi_i, \bar{\phi})]^2$
\end{itemize}

\begin{figure}[t!p]
\centering
        \subfloat{\centering\includegraphics[width=0.32\textwidth]{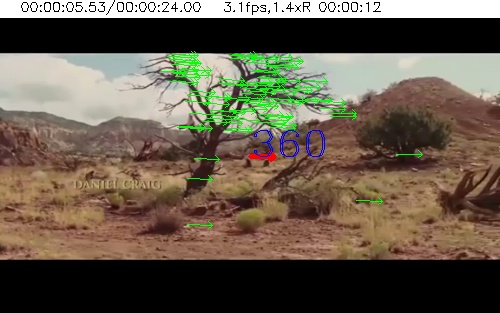}}
        \subfloat{\centering\includegraphics[width=0.32\textwidth]{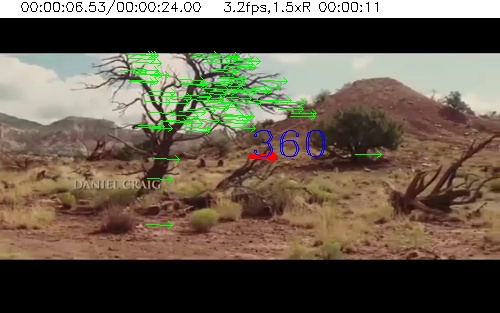}}
        \subfloat{\centering\includegraphics[width=0.32\textwidth]{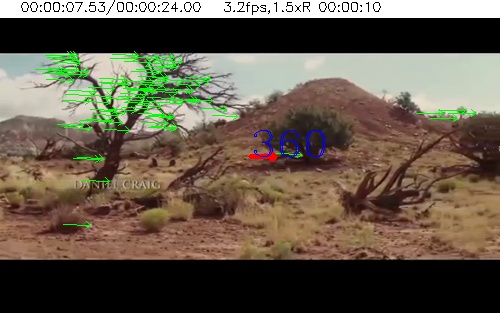}}
        \vspace{-1.2em}
        \subfloat{\centering\includegraphics[width=0.32\textwidth]{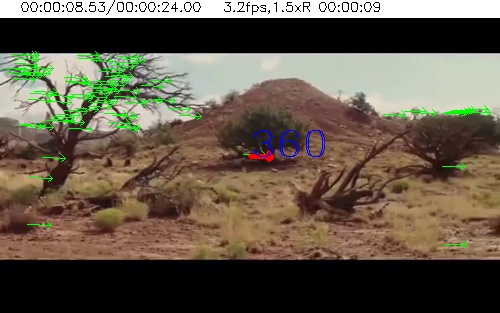}}
        \subfloat{\centering\includegraphics[width=0.32\textwidth]{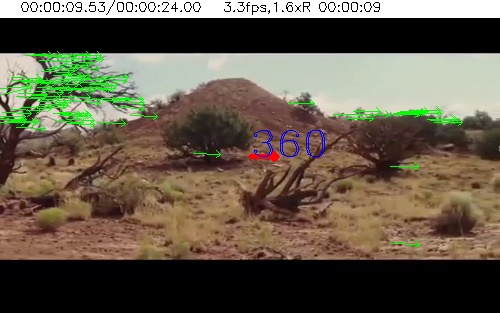}}
        \subfloat{\centering\includegraphics[width=0.32\textwidth]{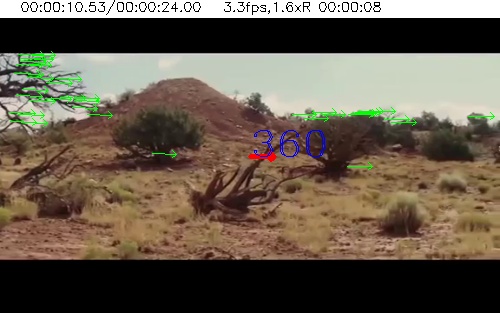}}
   
\caption{Example of computing the pan-tilt-pedestal-truck confidence movement feature. The frames are taken from a clear truck scene from the movie \textit{Cowboys and Aliens}.} \label{fig:panningExample}
\end{figure}

Figure \ref{fig:panningExample} presents an example of a panning scene from the \textit{Cowboys and Aliens} movie. The green vectors correspond to the extracted flow vectors and it is obvious that they share (almost) the same angle and relatively high magnitudes. Also, the corresponding PTPT confidence measure is 15 times higher than in other scenes in the movie not characterized as pan-tilt-pedestal-truck. Therefore, the adopted ratio is high for this example, as intended.
 
\subsubsection{Facial Information}
The existence of faces and the way they are illustrated are rather important characteristics in cinematography. Close-cuts to characters are often given to leading characters in films, in order to indicate their importance. In this work, we have have selected to apply the widely used Viola-Jones method in order to detect faces \citep{viola2004robust} in each frame of the film. Then, we calculate the following features related to faces: (a) the number of detected faces per frame and (b) the ratio of the face's bounding box area to the overall frame size. 

\subsubsection{Shot Length Information}
Film transition is an important procedure in cinematography applied in the post-production phase by combining shots and scenes. Shots are sequences of successive video frames that have been captured without interruption by a single camera. Shot transition is usually achieved through simple cuts, two different successive shots are played played one after another. Some directors are known for using ``long takes'', i.e. shots that last longer than usual. The film \textit{Rope} by Alfred Hitchcock is the first widely known movie that contained long takes. 

Shot change detection is a task that has been massively studied in video analysis \citep{cotsaces2006video, hanjalic2002shot}. Our goal in the context of this work was not to extract a fully accurate shot boundary detection estimate, but to calculate an aggregate measure of shot length, also taking into account the camera movement. Therefore, we have adopted three basic thresholding rules applied on (a) the number of significantly changed pixels between two successive grayscale frames (b) the overall motion based on the sum of magnitude of the flow vectors and (c) the sum of absolute differences of the gray value histograms between two successive frames.

\begin{figure}[ht]
    \centering
    \includegraphics[width = 0.8\textwidth]{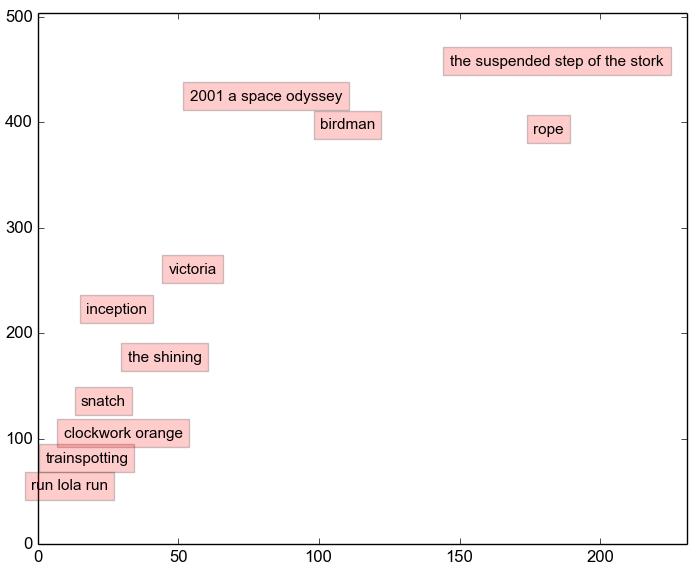}
    \caption{Differences in average shot lengths between movies with fast and slow camera movements. The x-axis denotes the average shot length and the y-axis the average top-10 values of the shot lengths.}
    \label{fig:demoshot}
\end{figure}

Figure \ref{fig:demoshot} illustrates the distribution of 11 movies in their average (x axis) and average top-10 values of their shot lengths (y axis). Movies like \textit{Run Lola Run} and \textit{Trainspotting} share very low average shot lengths and top $10\%$ average shot lengths, since these movies have very abrupt cuts and fast camera movements. Angelopoulos's movies (\textit{The Suspended Step of the Stork}) are known for their slightest movements and changes, as well as long takes. 

\subsection{Metadata Analysis}
Feature extraction from metadata is much more straightforward, since they only used as auxiliary information in our work, which focuses on low-level information. Utilizing publicly available information regarding the cast, the directors and the genres of the movies in our collection from \textit{IMDB}, we create a categorical vector for each movie, where each cell contains a binary value, $0$ or $1$, denoting relation between the movie and the corresponding tag. These tags are the different actors, directors and movie genres found in our collection. Due to the small number of movies the final representation has approximately $630$ unique features ($\approx500$ actors, $\approx110$ directors, $\approx20$ genres).  

\subsection{Content Similarity and Data Fusion}
Having represented the movies as feature vectors, we can define similarity between these vectors to correspond to the similarity of their respective movies. We compute the \textit{cosine similarity} between all movie pairs ($\vec{m_a}, \vec{m_b}$), in the different representation spaces:

\begin{equation}
 CosSim(\vec{m_a}, \vec{m_b}) = \frac{\vec{m_a}\times \vec{m_b}}{\|\vec{m_a}\|\times \|\vec{m_b}\|}   
\end{equation}

This results in a \textit{similarity matrix} between movies for each modality. In order to combine these content-specific similarities we adopted a simple weighting scheme between the similarity matrices, where the optimal weights for each modality are set after extensive experimentation.

\subsection{Computational Complexity and Implementation}\label{sub:scal}
 All of the methods described above have been implemented in the Python programming language, using open source libraries for data handling, computer vision, signal analysis and machine learning routines. Experiments have been conducted on a standard personal computer with an Intel(R) Core(TM) i7-5820K CPU @ 3.30GHz processor and 16GB of RAM. The computational time required for each modality, along with the computational complexity of the basic components of each modality, are described bellow:
 \begin{itemize}
    \item text mining and topic modeling is achieved in just $0.05\%$ of the real movie's duration. The theoretical complexity of LSI is  $\mathcal{O}(N^2*K^3$) where $N$ is the number of words in the collection and $K$ the number of topics. However, in recent years it has been reduced to the SVD level \citep{Ding:2011,Holmes:2007}. LDA with gibbs sampling on the other hand has a complexity of $\mathcal{O}(DKM)$ where $D$ is the number of documents-movies, $K$ number of topics and $M$ the mean document length. Still, there also exist techniques \citep{Porteous:2008, Newman:2008} for speeding up this procedure as well. Let us note here, that these procedures need to run only once over all the subtitle texts in our collection, as opposed to the per movie complexity described below for the audio-visual domain.
    \item audio and music analysis is performed in a $100 \times$ realtime rate, meaning that, on average, a movie's audio content is analyzed at a time equal to $1\%$ of its duration. The most computationally burdensome component of the audio analysis module is feature extraction, and in particular the spectral computation. Since FFT has been adopted to extract the spectral descriptors of the audio signal, this complexity is theoretically equivalent to
    $\mathcal{O}(n\log n)$ where $n$ is the number of samples per window. The duration of a movie is $D=n*N$ where $N$ is the window size, so the total complexity is $\mathcal{O}(D*\log n)$.
    \item visual feature extraction is more computationally demanding, and as described above, we only require to analyze 2 frames per second. Given that, on average the analysis is performed in a $6 \times$ realtime rate (e.g. it takes almost 17 minutes to analyze a movie of 2 hours). So the analysis of the visual domain is achieved, on average, in $17\%$ of the real movie's duration. For the visual domain the flow extraction is the most burdensome and it has been proven \citep{Lucas:2004} to be $\mathcal{O}(n^2*N+n^3)$ where $n$ is a constant parameter of the algorithm and $N$ the total number of pixels. So the final complexity is linear to the number of pixels $N$ (and obviously linear to duration $D$) resulting in $\mathcal{O}(D*N)$ complexity for the analysis of each movie.
 \end{itemize}
 
 To sum up, in order to extract the whole set of features from a movie, almost $18\%$ of its real duration is required. This means that for a product-case dataset of, say, 10,000 movies the computational time required is almost 23 days, if all CPU kernels of a i7 processor are used on a single computer. Equivalently, just 4 VMs would be required to compile the respective dataset of content movie similarities in less than a week.

\section{Dataset}\label{sec:data_gt}
\subsection{Data Description}
In order to prove the ability of the low-level modalities to improve the performance of the content similarity, compared to the metadata information, we have compiled a real-world dataset of 160 movies. These movies have been selected from the \textit{Top 250 Movies}\footnote{\url{http://www.imdb.com/chart/top}}. Our purpose was to use movies that are widely known and therefore the quality of the results can be easily assessed. Moreover, the dataset is populated with different types of movies to avoid metadata-specific bias, such as genre or casting. The subtitles were downloaded from an open source database\footnote{\url{http://www.opensubtitles.org/en/search}} and were hand-checked for mistakes. 

\subsection{Ground-truth Generation}
In order to evaluate the similarity rankings generated by the different modalities, we need a \textit{ground-truth} similarity between the movies of the dataset, against which we can pitch our results. Towards this end, we used the \textit{Tag-Genome} \citep{vig:tag} dataset to create a ground-truth similarity matrix between the movies. Every movie is represented as a vector in a tag-space with $\approx1100$ unique tags. The tags can be a wide variety of words-phrases such as adjectives (``funny'', ``dark'', ``adopted from book''), nouns (``plane'', ``fight''), metadata (``tarantino'', ``oscar'') and so on, that act as descriptors for the movies. Having this representation for each movie we obtained the ground-truth movie similarity matrix, through calculation of the cosine similarity metric between each pair of movies, as already mentioned.

\section{Experimental Results}\label{sec:res}
In the context of experimental setup, our goal was to evaluate the performance of the low-level modalities (when used either individually or in a fusion approach)  in terms of content similarity and knowledge discovery. In this Section we provide the following types of experimental results: (a) a qualitative evaluation based on recommendation metrics and (b) a use case on how particular low-level features achieve differentiation between directors and movie genres and (c) a movie network that demonstrates the usefulness of our approach.

\subsection{Recommendation Evaluation}
Firstly, in order to rank the quality of the similarities for each model, we utilized the similarity rankings created by the aforementioned matrices. Specifically, for each movie we are interested in the similarity ranking of the first two recommendations generated by each model. We calculate the median position, of the first and second recommendations over all movies, as ranked in the ground truth similarity matrix. This information-retrieval measure conveys the similarity ranking accuracy for each model. Moreover, in order to estimate a proportion of the ``good'' recommendations each model provides , we also calculate a recall type of measure. Specifically, it is the percentage of recommendations, averaged over all movies, that are in the Top 10 most similar movies, according to the ground truth similarity ranking for each movie. This indicates the sensitivity of each model. 

Tables \ref{tab:sing}, \ref{tab:fus_md} and \ref{tab:fus_int} present the results for each individual model, the fusion of each model with the metadata model and finally the fusion of specific models. Regarding Table \ref{tab:sing}, we can see that the metadata-based model retrieves the best recommendations. This is only natural, because the manual tags of the ground-truth data are essentially a super-set of the metadata and semantically much more similar to them, than the features generated from the rest of the models. Concerning the content based models, the best ones are the subtitle-based models and especially the \textit{LDA} model, but with minor differences to the other textual models. Finally, from the low-level audio-visual models, video is by far the best of the three, followed by the audio and in the end the music model.

\begin{table}[!htb]
      \centering
        \begin{tabular}
        {M{0.1\textwidth}M{0.19\textwidth}M{0.12\textwidth}M{0.12\textwidth}M{0.12\textwidth}M{0.12\textwidth}}\hline
        Modality & Model & Median Ranking 1st Rec & Top 10$\%$ of 1st Rec & Median Ranking 2nd Rec & Top 10$\%$ of 2nd Rec \\ \hline
        \multirow{3}{0.1\textwidth}{Text (subtitles)} & tf-idf & 18.0 & 40.0 & 26.5 & 28.1\\ 
        & LSI & 17.0 & 41.3 & 22.5 & 33.0\\
        & LDA & 15.5 & 43.0 & 24.0 &	33.75 \\  \hline
        \multirow{3}{0.1\textwidth}{Audio visual} & Music (M) &  55.5 &	8.8  & 61.0 & 10.0\\ 
        & Audio (A) & 51.0 & 11.9 & 53.0 & 10.6\\ 
        & Video (V) & 47.0 &	20.6 &	42.0 &	16.9\\ \hline
        Metadata & Metadata (MD) & 8.0 & 55.6 &9.0 & 53.1\\ \hline
        \end{tabular}
        \caption{Performance measures for individual modalities. As expected, the metadata-based content similarity achieves the best performance, since it is based on manually-provided content tags.}\label{tab:sing}
\end{table}

In Table \ref{tab:fus_md}, we can see the performance measures of all the individual models, \textit{fused with the metadata model}. The values beside each model are the optimal weights for the specific fusion. Moreover, we are also reporting here and in Table \ref{tab:fus_int} the cases where the results are statistically significant better than the standalone metadata model. The test results stem from the \textit{Wilcoxon signed-rank test} \citep{Wilcoxon:1945}, which is a non-parametric hypothesis test for paired samples. It is the equivalent of the paired \textit{paired Student's t-test} but without the assumption of normal distribution of the samples. The Wilcoxon signed-rank test is chosen because we want to study the effects of recommendation over a set of movies, so the per movie recommendations should be paired between two different models. Also, the larger the difference between the rankings of the pair (between the two models on a specific movie that is), the more weight it gains in the test, utilizing the differences both in \textit{direction} and \textit{mangitude} for the per-movie recommendations pair \citep{Siegal:1956}.

Examining the table, the most important conclusion is that \textbf{fusing with any of the low-level modalities boosts the performance of the metadata model, for the 1st recommendation ranking, almost by $50\%$}. This is probably the most interesting outcome of this experimentation, considering that the individual models perform worse than the metadata model, consequently meaning that there is much diversity in the recommendations given by the individual modalities and the metadata model, for most of the movies. In more detail,  fusing with the subtitle-based models gives better results, but that is not always the case for all the measures,  (e.g.  fusion with the video model performs better in the top 10$\%$ of 1st recommendation measure). Moreover, it is important that even fusing with the less accurate individual models, such as audio and music, still gives us a boost in performance over the standalone metadata model. A final interesting note, is that in the case of fusing audio and metadata, the similarity matrix of audio plays the most important role in the fusion with a $70\%$ weight factor, as opposed to the rest fusion weights where the metadata matrix has the highest weighting term. 

\begin{table}[!htb]
        \centering
        \begin{tabular}{M{0.25\textwidth}M{0.13\textwidth}M{0.13\textwidth}M{0.13\textwidth}M{0.13\textwidth}}\hline
        Model & Median Ranking 1st Rec & Top 10$\%$ of 1st Rec & Median Ranking 2nd Rec & Top 10$\%$ of 2nd Rec \\ \hline
        tf-idf (0.28) & \textbf{3.0}\textsuperscript{**}	& 63.8\textsuperscript{***}  & 9.5 &50.0\\ 
        LSI (0.18) & \textbf{3.0}\textsuperscript{**} & 64.4\textsuperscript{***}  &\textbf{9.0} & \textbf{53.1}\\
        LDA (0.11)& 4.5\textsuperscript{-} &60.0\textsuperscript{-}  &10.5 &48.1 \\  \hline
        Music (0.14) &5.0\textsuperscript{-} &61.3\textsuperscript{*}  &9.5 &50.0\\ 
        Audio (0.70) & 5.0\textsuperscript{***} &57.5\textsuperscript{-}  &12.0 &45.6\\ 
        Video (0.13) & 4.0\textsuperscript{-}  &\textbf{66.3}\textsuperscript{***} & 10.5 & 48.8\\ 
        \bottomrule
          \multicolumn{5}{c}{\rule{0pt}{ 1.2\normalbaselineskip}% strut
            Statistical significance levels \textsuperscript{***}$p < 0.01$, 
            \textsuperscript{**}$p < 0.05$, 
            \textsuperscript{*}$p < 0.1$,} \textsuperscript{-}$p > 0.1$
                \end{tabular}
        \caption{Performance measures for all individual modalities \textit{fused} with metadata. In all cases, the fusion performance is boosted compared to the metadata accuracy, almost $50\%$. }\label{tab:fus_md}
        
\end{table}

Finally, we also tried to fuse more than two models together and some of the best combinations are presented in Table \ref{tab:fus_int}. We did not discover any boost in the performance of the fusion models presented in Table \ref{tab:fus_md}, nor a significant decrease over the best performing model from Table \ref{tab:fus_md}.

\begin{table}[!htb]
        \centering
        
        \begin{tabular}{M{0.29\textwidth}M{0.13\textwidth}M{0.13\textwidth}M{0.14\textwidth}M{0.13\textwidth}}\hline
        Model & Median Ranking 1st Rec & Top 10$\%$ of 1st Rec & Median Ranking 2nd Rec & Top 10$\%$ of 2nd Rec \\ \hline
        M, A, V \\  0.06, 0.84, 0.1 & 32.0	&25.0 & 51.0 &15\\ \hline
        LSI, M, A, V \\0.06, 0.62, 0.12, 0.20 &10.0 &    48.1 & 17.0 & 31.9 \\\hline
        M, A, V, MD\\ 0.13, 0.17, 0.05, 0.56 & \textbf{4.0}\textsuperscript{-} &  \textbf{66.3}\textsuperscript{***}  &  11.5 &  41.3 \\\hline
        LSI, M, A, V, MD\\ 0.12, 0.06, 0.13, 0.1, 0.59 & \textbf{4.0}\textsuperscript{**} &65.6\textsuperscript{***}  &\textbf{9.5} &\textbf{50.0}\\
        \bottomrule
          \multicolumn{5}{c}{\rule{0pt}{ 1.2\normalbaselineskip}% strut
            Statistical significance levels \textsuperscript{***}$p < 0.01$, 
            \textsuperscript{**}$p < 0.05$, 
            \textsuperscript{*}$p < 0.1$,} \textsuperscript{-}$p > 0.1$
                \end{tabular}
        \caption{Performance measures for fusion between (a) all low-level modalities (audio, music, video) (b) all low-level modalities with text (c) all low-level modalities (audio,music,video) with metadata and (d) all content modalities (audio,music,video,text) with metadata }\label{tab:fus_int}
\end{table}

\subsection{Modality Features Differentiation per Genre and Director}

In order to gain more insight, regarding the discrimination capabilities offered by each individual model, their complementarity with the metadata model and their possible specialization in particular types of movies, we also implemented two simple genre and director specific tasks. 
Specifically, we firstly group the movies, \textit{per genre} and according to the \textit{director} of the film. Then, for each movie in our collection we retrieved the recommendations from the fusion models in Table \ref{tab:fus_md}. Based on the returned recommended movies, we examined which belonged to the same genre (or director) as the query movie. Finally we computed the average ratio of relevant to the total number of retrieved movies (with regards to either genre or director).

The core idea of these tasks, is to gain an initial intuition regarding some very interesting questions about the way humans perceive similarity between movies and what different modalities are involved in this perception. This is paramount, in order for us to gain insights about whether there are specific modalities more capable of dealing with specific genres, or whether some directors can be identified based on specific low-level features provided by a modality etc.

\begin{figure}[ht]
    \centering
    \includegraphics[width = 1\textwidth]{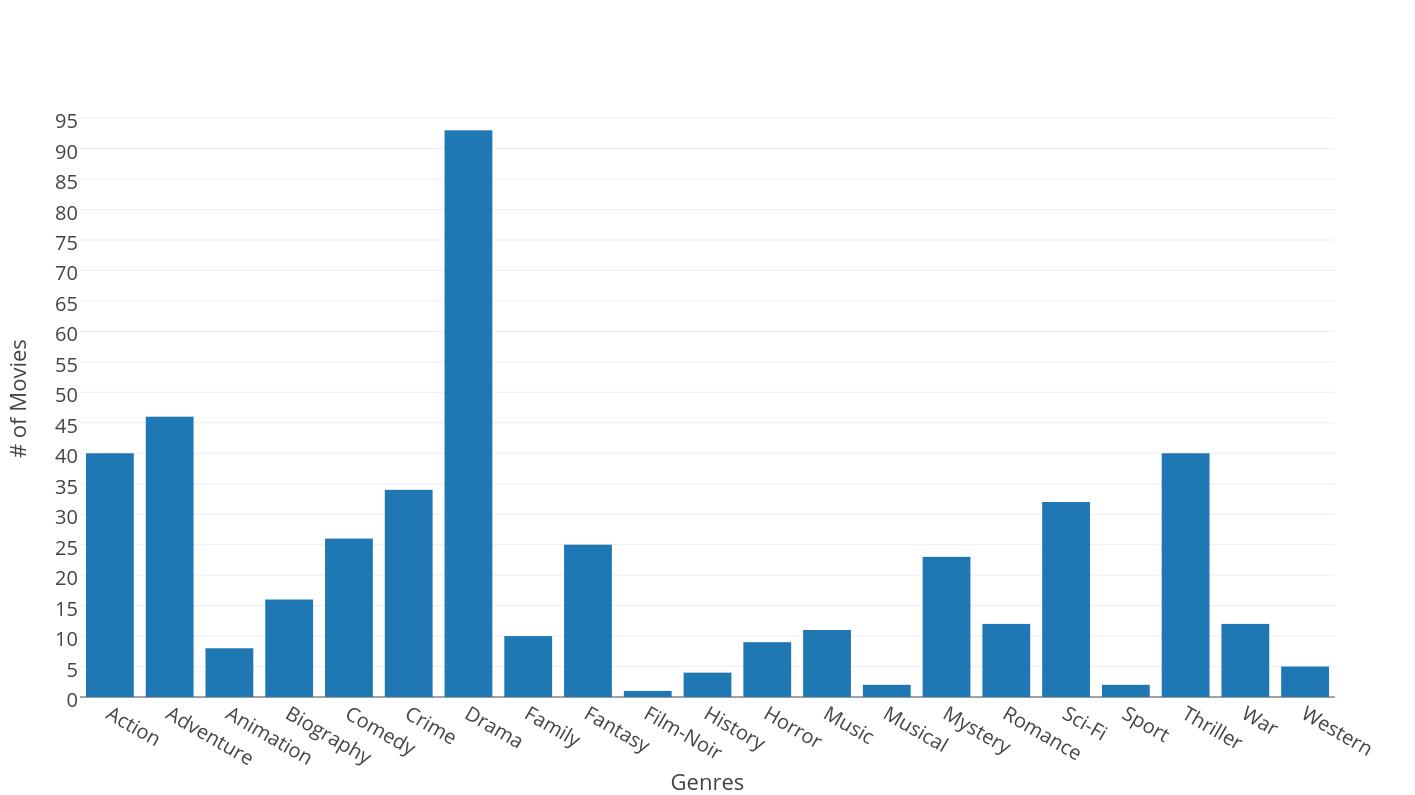}
    \caption{Distribution of movies in the different genres.}
    \label{fig:genre_distr}
\end{figure}

\begin{figure}[ht]
    \centering
    \includegraphics[width = 1\textwidth]{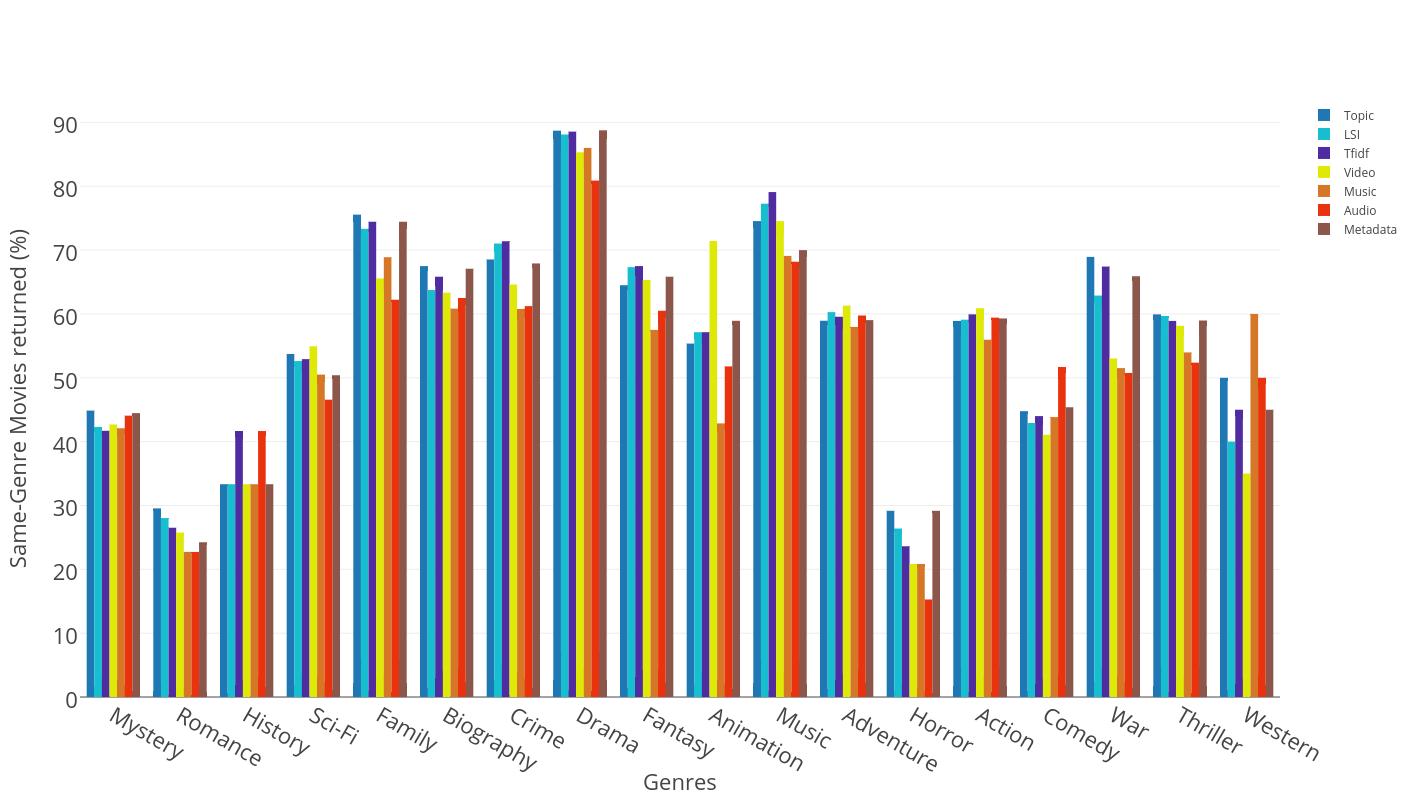}
    \caption{Averaged percentage of same-genre recommendations across all individual models fused with the metadata model. }
    \label{fig:genre_discr}
\end{figure}

Before proceeding with the results, since we mainly focused on the per genre grouping of the movies task, the following analysis will be heavily centered around that task. In Figure \ref{fig:genre_distr} we see the distribution of movies in the different genres found in our collection. There are in total $21$ different genres and each movie could belong in more than one genre, which is also the most common case. The results are shown for all models and genres in Figure \ref{fig:genre_discr}. For each different genre and model, we can see the percentage of recommendations that belonged to the same genre as the queried movie, averaged across all movies. As a reminder, the individual models are fused with the metadata according to the weights of Table \ref{tab:fus_md}. The first 3 bars, colored in shades of \textit{blue}, in each genre, are the models based on the textual modalites, followed by the video model in \textit{yellow} and the music (\textit{orange}), audio (\textit{red}) models next and finally the individual metadata model (\textit{brown}). We have omitted \textit{Film Noir, Musical} and \textit{Sport} genres, because we did not have enough movies to get a sound measurement of the same-genre retrieval ratio. Examining the figure leads to many interesting observations:
\begin{enumerate}
\item Fusion with \underline{textual} models outperforms the individual metadata model  in \textit{Romance} (All models), \textit{History} (LSI), \textit{Sci-fi} (LDA) and \textit{Actionc} (tf-idf) among others. This substantiates the logical intuition that same-genre movies can be thematically connected and may exhibit the same vocabulary more or less, for specific genres; \textit{Romance} would be an obvious example, as  all textual fusion models outperform the metadata model. 
\item The \underline{video} fusion model outperforms the metadata model in many genres, such as \textit{Sci-fi, Animation, Adventure} and others. \textit{Animation} is a striking example one could think of, where information from visual features could be indispensable in order to find similar movies due to their particularities. \textit{Adventure} and \textit{Action} are other such cases where the fast transition of scenes and flow of movement can be identified using visual cues.
\item The \underline{music} fusion model achieves noticeably better results in \textit{Western} genre, probably due to the idiosyncratic musical pieces used in such films. The same stands for the \underline{audio} fusion model for this genre, as well as \textit{History} and \textit{Comedy} films among others.
\end{enumerate}

Ultimately, the subtitles-based fusion models outperform the metadata model in $13$ out of $21$ genres, the video fusion model in $6$ genres, while the sound-based models, music and audio, perform better in $2$ and $4$ genres respectively.

Regarding the director task, we followed the same methodology focusing on the top-10 most prolific directors, regarding movie population as distributed in our collection, in order to have more robust results. However, with very few exceptions the results were not indicative of significant differentiations between group of movies directed by the same person through the fusion of different models. A brief summary of those exceptions, where the fused models perform better that the metadata model, would be as follows:

\begin{itemize}
\item \underline{Textual} fusion models perform better on movies directed from \textit{Alfred Hitchcock} and \textit{Quentin Tarantino} than the individual metadata model. 
\item The \underline{visual} fusion model performs better with movies by \textit{Quentin Tarantino} and \textit{Steven Spielberg}.
\item Finally, \underline{music} fusion model in movies by \textit{Kubrick} while the \underline{audio} fusion model produces better results in movies by \textit{Robert Zemeckis}. 
\end{itemize}

However, the above results should be taken with the grain of salt because of the small amount of movies by each director, the distribution of those movies over genres that could affect the results (we may be essentially differentiating between genres, while thinking we differentiate between directors because of the genre-director correlation) and the small differences in performance between the fusion and individual models (further accentuated by the limited dataset).

\subsection{Movie Network Demo}

\begin{figure}[ht]
    \centering
    \includegraphics[width = 1\textwidth]{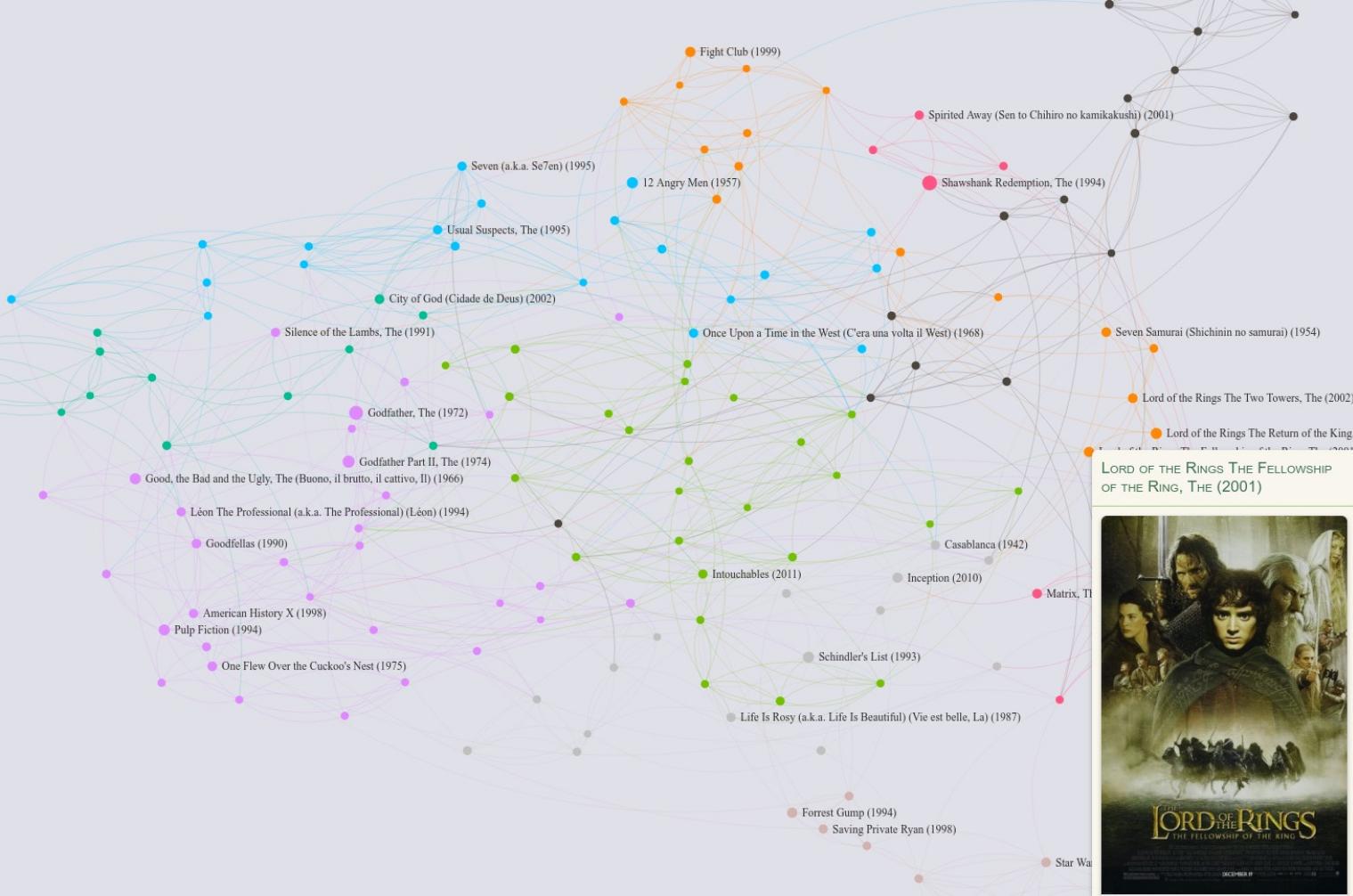}
    \caption{Instance of the Movie Similarity Network Demo. Nodes are movies, links from each node represent similarity and the color of the node is according to the community it belongs to}
    \label{fig:movie_network}
\end{figure}

Finally, in order to demonstrate the usefulness of our approach with the multimodal representations of the movies we offer an online interactive demo\footnote{\url{http://users.iit.demokritos.gr/~bogas.ko/movies/examples/movies_network.html}}, where the movies have been represented as a network graph.  A static screenshot of the demo can be seen in Figure \ref{fig:movie_network}.

Each node is a movie from our dataset and the links between the movies denote the similarities as found by our models. The node size corresponds to the score of the movie according to \textit{IMDB}\footnote{\url{http://www.imdb.com/}} and the node is colored according to which community it belongs to. The communities in the graph are found using the \textit{Louvain method} based on modularity maximization \citep{blondel:louvain}. It is very interesting to see the differences in communities found in these similarity networks produced by the different modalities. The user can use different kind of filters, based on metadata, content and communities as found by each representation, in order to gain insights about movie relations through different aspects. This per model movie similarity browsing can give us insights about the differentiations and strengths of each modality, as for example one can see a cluster of movies in the video model that mainly consists of black-and-white films (dark brown community).

\subsection{Discussion}
In this paper, we examined the ability of all modalities (textual, auditory and visual) involved in a cinematic movie to provide content similarity measures, in the context of a movie recommendation application. Moreover, we validated the basic premise of our research regarding the latent connection between low-level features of movies and human-level semantics of similarity.

In more detail, we've shown the usefulness of raw features when fused together with metadata to provide better recommendation of movies. Regarding the \textit{textual} models, it has been proven that the \textit{LDA} performs better than \textit{LSI} and \textit{tf-idf}. Moreover, \textit{LDA} offers us a topical representation of the movies that is much closer to human understanding and allows for better interaction and browsing of the movies.  However, when fused with metadata or other models \textit{LSI} seems to be the best choice, which is also much less time consuming than \textit{LDA}.

Regarding the \textit{audio}-\textit{visual} domain, supervised and unsupervised methods for extracting information have also been proposed. In particular, in the audio domain pretrained classifiers have been used to extract statistics about the existence of particular audio classes and musical genres. In the visual domain, we have extracted some characteristics that are associated with particular filmmaking techniques (e.g. camera movement, shot lengths, etc). Experiments have proven that \textit{the visual cue is more informative with respect to the movie content similarity than audio and music}.

We've also shown the differentiation of results when using different data modalities in a variety of tasks. As noted, the visual component of a movie can be of great value when searching for movies based on specific cinematographic techniques. Also, the audio-music elements of the movie can be telling in terms of the genre and the thematic of the movie.

This novel way of representing movies, as multimodal data sources, opens up new horizons in the ways we interact with movies, allowing as to tap into the latent knowledge found in these representations. This can pave the path for more holistic approaches in movies recommendation, as showcased in the demo of the network of movies, and address a series of problems in recommender systems. For example, the presented approach would be invaluable when dealing with the \textit{cold start} problem \citep{Schein:2002} and recently premiered movies must be associated with existing ones, because this approach is not bounded by the perception of similarity and the ratings of the users. 

However, the presented work also has some limitations. Firstly, the results presented here are only based on a small dataset of movies. In order to further validate the insights gained from this work, experiments on a larger scale should be conducted. This would also help defining a strict cross-validation scheme and experiment with the efficiency of this approach when addressing other important matters of recommender systems \citep{Khusro2016}. Also, more consideration should be put on the way the ground truth similarity is generated. Firstly, it is conceptually related to the metadata features, unfavorably favoring the metadata model when comparing results. Moreover, it is not trivial to define actual similarity between movies. Do ratings of movies generated by users implicitly express this similarity? Or should an explicit process of manual linking of similar movies take place? This is a very important matter, as it heavily affects the experimental results.

\section{Conclusions and Future Work}\label{sec:fut}
 
The results presented in this paper verify the core idea of adopting multimodal information to boost the performance of movie recommendation systems. The basic outcomes of our research are the following:

\begin{enumerate}
    \item The most important and promising outcome of the experimentation is that \textbf{low-level feature models exploit latent information that boosts the performance of human-generated information models (metadata) at almost a $50\%$ ratio, despite the fact that their individual performances are much lower.} This implies that the diversity between the decisions from different modalities is high. These results prove that the proposed low-level features can be adopted in the context of a multimodal content-based recommendation system. However, fusing \textbf{all} modalities did not further improve the performance of the content similarity approach: this demonstrates that in future work, combining more than two modalities should be handled using more sophisticated approaches (some ideas are presented in detail below)
    \item The workflow for a complete methodology for automatic similarity extraction for movies based on low-level features has been described and evaluated. Detailed examples on how modality-specific features discriminate between different cinematographic attributes are presented. In addition, a detailed experimentation on content similarity has been conducted, based on a dataset of 160 movies. 
    \item Finally, we have showcased examples where specific modalities seem to be good at differentiating between separate genres of movies, and to a lesser extend different directors. This implies that different modalities and, maybe, specifically a selection of low-level features from those modalities, can capture high-level concepts of similarity or cinematographic styles, as defined by humans.
\end{enumerate}

At the same time, they inspire several future (and ongoing) research directions. In particular:

\begin{itemize}
\item Scalability: We have already discussed that the proposed approaches require a proportion (less than $\frac{1}{5}$) of the real movie duration to extract multimodal knowledge. They can be therefore applied to larger datasets in order to simulate their ability to perform on real movie recommendation systems. 
\item With regards to the particular low-level modality analysis methods:
    \begin{itemize}
        \item In the \textit{text} analysis module, other topic methods could be evaluated such as Hierarchical Dirichlet Processes \citep{teh:hdp}.
        \item Regarding the \textit{audio} domain, more detailed class representations will be added as pretrained supervised models, in order to cover more audio classes and musical genres. However, our goal in this task is to also include unsupervised similarity extraction methods, that discover content similarities based on clustering of the audio feature distributions. Additionally, temporal methods (e.g. HMM or LSTM approaches) will be used to also model the way features and classes change over time.
        \item Similarly, in the \textit{visual} domain unsupervised similarity extraction and temporal modeling will be adopted. Additionally, we will focus on extracting higher level information from all visual cues. Particularly, regarding face-related analytics, we are already building methods that discriminate between different types of close cuts (medium, extreme, lean-ins, etc). Also, face clustering will also be implemented to achieve a more detailed representation of the existence of faces, so that the features will answer questions like: how many faces appear in the movie or which are the most dominant faces. Regarding the camera movement features, we will generalize the existing method to also discriminate between different types of camera movement (panning, zooming, truck, etc), leading to more detailed high-level and distinctive filmmaking styles. Finally, more accurate shot length extraction are being implemented, while we also focus on achieving classification between different types of shot transitions: simple cuts, wipes, fade-ins, fade-outs, etc.
    \end{itemize} 
\item Regarding multimodal fusion we are already examining more sophisticated fusion schemes that also take into consideration temporal dependencies and correlations between the different modalities.
\item Finally, in order to achieve a fully functional and complete recommendation system, knowledge with added \textit{user preferences} will be included by adopting collaborative filtering methodologies. In addition, user clustering and profiling information will be correlated with low-level knowledge from multimodal information, in order to discover if different groups of user preferences correlate better with different modalities (e.g. if certain people choose the movies based on the visual filmmaking characteristics or the topic, etc).
\end{itemize}

%
% The following two commands are all you need in the
% initial runs of your .tex file to
% produce the bibliography for the citations in your paper.
\bibliographystyle{model5-names}
\biboptions{authoryear}
\bibliography{bibl}  % sigproc.bib is the name of the Bibliography in this case

%\begin{acknowledgements}
%If you'd like to thank anyone, place your comments here
%and remove the percent signs.
%\end{acknowledgements}

% BibTeX users please use one of
%\bibliographystyle{spbasic}      % basic style, author-year citations
%\bibliographystyle{spmpsci}      % mathematics and physical sciences
%\bibliographystyle{spphys}       % APS-like style for physics
%\bibliography{}   % name your BibTeX data base

\end{document}